\documentclass[aps,prb,twocolumn,showpacs,floatfix,groupedaddress,superscriptaddress]{revtex4-2}
\pdfoutput=1
\usepackage[linkcolor=blue,colorlinks=true,breaklinks=true,citecolor=blue,urlcolor=blue]{hyperref}
\usepackage{graphicx,graphics}  
\usepackage{dcolumn}   
\usepackage{braket}
\usepackage[mathscr]{euscript}
\usepackage{amsbsy}
\usepackage{bm}        
\usepackage{amssymb}   
\usepackage{multirow,tabularx}
\usepackage{physics,amsmath}
\usepackage{dsfont}
\usepackage{xcolor}
\usepackage{comment}
\usepackage{dirtytalk}
\usepackage[titletoc,title]{appendix}
\usepackage[dotinlabels]{titletoc}
\usepackage[caption=false,labelformat=simple]{subfig}

\begin{document}

\title{Orbital Hall Conductivity in a Graphene/Haldane and Haldane/Haldane Bilayers }
\author{Sovan Ghosh}
\email{sovanghosh2014@gmail.com}
\affiliation{Department of Physical Sciences, Indian Institute of Science Education and Research Kolkata\\ Mohanpur-741246, West Bengal, India}
\author{Bheema Lingam Chittari}
\email{bheemalingam@iiserkol.ac.in}
\affiliation{Department of Physical Sciences, Indian Institute of Science Education and Research Kolkata\\ Mohanpur-741246, West Bengal, India}
\begin{abstract}
We investigate the orbital Hall conductivity in bilayer graphene (G/G) by modifying one or both the layers as Haldane type ($\rm  G/ \tilde G$ : Graphene/Haldane  and $\rm \tilde G/ \tilde G$ : Haldane/Haldane) with the inclusion of next nearest neighbour (NNN) hopping strength ($t_2$) and flux ($\phi$). It is observed that the low energy bands of $\rm  G/ \tilde G$ and $\rm \tilde G/ \tilde G$ are isolated with a gap at charge neutrality with the next nearest neighbour (NNN) hopping term $t_2e^{\pm i\phi}$. The time reversal (\textit{TR}) symmetry breaking with $t_2e^{\pm i\phi}$ induces a large orbital magnetic moment ($\vb{m}_n(\vb{k})$) for the $n^{th}$ band in $\rm  G/ \tilde G$ and $\rm \tilde G/ \tilde G$ bilayers. This \textit{TR} symmetry breaking, modulated by the $t_2$ strength, leads to the emergence of {\it Orbital Ferromagnetism} and {\it Valley Orbital Magnetism} within the BZ for the Haldane single layer as well for both $\rm  G/ \tilde G$ and $\rm \tilde G/ \tilde G$. We show that for the applied longitudinal electric fields, the intrinsic angular momentum ($L^z$) gives the orbital current ($\mathcal{J}^{z,orb}$) along a transverse direction and generates the orbital Hall conductivity (OHC). We further show that the orbital magnetic polarity leads the Haldane single layer to {\it Orbital Chern Insulator} with the quantized OHC in the gap over the occupied bands. Moreover, the accumulation of orbital magnetic moment of the bands in Haldane graphene bilayer shows {\it Quantum Orbital Hall Insulator} and {\it Orbital Chern Insulators}. Similarly, we show that in the hetero-bilayers, one of the layers of the Haldane type generates the orbital magnetism and induces the OHC. We conclude that the isolated bands in Haldane graphene bilayers with external stimuli are of an orbital nature and have various quantum orbital Hall phases.
\end{abstract}
\maketitle
\section{\label{sec:1}Introduction}
\begin{figure}[h]
     \centering     
     \includegraphics[width=\columnwidth]{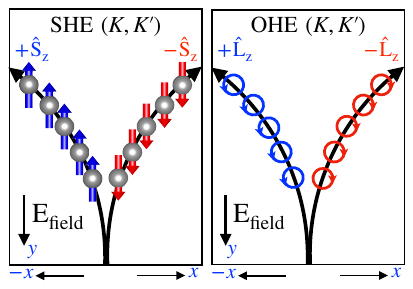}
     \caption{ Schematic diagram of spin Hall effect (SHE) and orbital Hall effect (OHE), where opposite spin ($\pm \hat{S}_z$) and orbital angular momentum ($\pm \hat{L}_z$) moving in two different edges (x direction) in the response of a longitudinal electric field (along y axis)}
    \label{fig:Schematic}
 \end{figure}

Two-dimensional quantum materials consist of single or multiple layers, including marginally aligned, commensurate stacked heterostructures  \cite{2Dquat,Andrei2020,Huang2020}. In these materials, spontaneous symmetry breaking due to external stimuli, such as magnetic or electric fields can lead to isolated low-energy bands and strong localization  \cite{Liu2021, Yankowitz2018, Kim.acs.nanolett.8b03423, PhysRevLett.122.016401, Chen2019, PhysRevB.99.235417, PhysRevB.102.035411, Sinha2020, PhysRevB.103.075423, PhysRevB.103.165112, PhysRevB.104.045413, PhysRevB.105.245124, CHEBROLU2023115526, PhysRevB.108.155406}. Interestingly, isolated low-energy bands formed by geometrical interfaces are often topological and exhibit orbital ferromagnetism  \cite{Sharpe.science.aaw3780,Sharpe.acs.nanolett.1c00696}. Furthermore, the interplay between the electron correlations and topology in these materials can give rise to various emergent quantum phases. Recent examples include multilayer graphene aligned on hexagonal boron nitride and magic-angle twisted bilayer graphene (tBLG), which exhibit spontaneous orbital magnetism and valley polarization induced by electron-correlation effects in low-energy bands \cite{Liu2021,Geisenhof2021,He2020,Grover2022}.  These isolated bands represent highly localized atomic orbitals that often contain unpaired electrons with no spin polarization in the absence of external fields. These electrons exhibit self-rotation within their orbitals, either clockwise or counterclockwise \cite{ARYASETIAWAN201987}, resulting in a net orbital angular momentum (OAM). The specific direction of self-rotation of these unpaired electrons can spontaneously break the symmetry under an external electric field, leading to the orbital Hall effect  \cite{PhysRevLett.121.086602,PhysRevB.98.214405}. The OHE has been observed in centrosymmetric systems using the inter-band Bloch basis without breaking symmetries~\cite{PhysRevLett.95.066601,PhysRevLett.123.236403,PhysRevLett.121.086602,PhysRevB.101.121112,PhysRevB.103.085113,RevModPhys.82.1959,PhysRevLett.125.227702,PhysRevLett.99.236809,PhysRevB.103.224426}. Examples include unbiased transition metal dichalcogenide (TMD) bilayers, where the nearly degenerate Bloch bands from the individual layers contribute significantly to the OHE~\cite{PhysRevB.105.195421}. In contrast, gapped graphene does not exhibit inter-band contributions to the Bloch basis, and the OHE arises solely from broken symmetries~\cite{PhysRevB.103.195309}. Similar to spin in spintronics, the orbital degree of freedom of carriers in isolated bands is actually involved in transport and storage \cite{RevModPhys.76.323,RevModPhys.80.1517,RevModPhys.87.1213}. In the presence of an electric field and due to spin-orbit interactions, opposite spins can diffuse in opposite directions, leading to a transverse spin current even in non-magnetic materials. This phenomenon is known as the intrinsic spin Hall effect (SHE). Fig.\ref{fig:Schematic} compares the Spin Hall Effect (SHE) with the Orbital Hall Effect (OHE). In response to an applied longitudinal electric field a transverse orbital current induced. 
This is due to the accumulation of opposite intrinsic orbital magnetic moments at different edges \cite{Go_2021, PhysRevB.101.121112}. Importantly, unlike SHE, spin-orbit interactions do not play a role in OHE \cite{PhysRevLett.95.066601,PhysRevLett.121.086602}. 
Meanwhile, the Anomalous Hall Effect (AHE) can occur in non-magnetic materials with broken time-reversal symmetry. In contrast, the Orbital Hall Effect (OHE) is a distinct phenomenon specific to materials lacking a net spin moment.  OHE arises due to broken time-reversal symmetry, similar to AHE, but the observed Hall conductivity stems from the orbital current. Motivated by recent advances in orbital magnetism, this study investigates the behavior of orbital magnetization in graphene and bilayer graphene with varying stacking configurations. Subsequently, we extended our analysis to other related heterostructures. The remainder of this paper is organized as follows. Section~\ref{sec:2} introduces the theoretical framework, including the tight-binding Hamiltonian formalism and modern theory describing the Hall conductivity in bilayer systems. Section~\ref{sec:3} presents  the results, band calculations, orbital magnetic moments, and orbital Hall effect for the Haldane graphene bilayers. Finally, Section~\ref{sec:4} concludes the study. 
\section{\label{sec:2}Methodology}
\subsection{Tight-binding Hamiltonian}\label{TB_hamil}
\begin{figure}[h]
\centering
\includegraphics[width=\columnwidth]{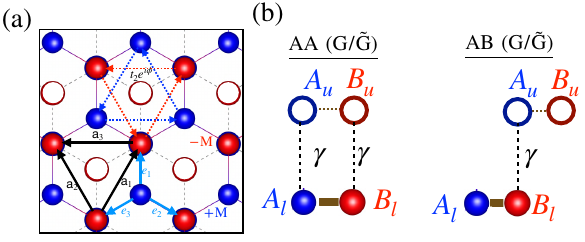}
\caption{(a) Schematic diagram of an AB-stacked bilayer graphene viewed from above. Here, $A_{l (u)}$ (blue solid spheres and circles) and $B_{l (u)}$ (red solid spheres and circles) represent the lower (upper) layer sublattices. The on-site energies on A and B sublattices are represented by $+M$ and $-M$, respectively. Vectors $\mathbf{e_1}$, $\mathbf{e_2}$, and $\mathbf{e_3}$ represent nearest-neighbor (NN) hopping, and vectors $\mathbf{a_1}$, $\mathbf{a_2}$, and $\mathbf{a_3}$ represent next nearest neighbor (NNN) hopping. The NN hopping strength is $t_g$, and the NNN hopping strength with flux is represented by $t_2e^{i\phi}$  (anticlockwise) and $t_2e^{-i\phi}$(clockwise). (b) Cross-sectional view of AA and AB stacked bilayer graphene with inter-layer hopping strength $\gamma$ }
\label{fig:schem}
\end{figure}

\noindent
We constructed a tight-binding Hamiltonian for bilayer graphene using maximally localized Wannier functions \cite{PhysRevB.56.12847, RevModPhys.84.1419}. The Wannier approach provides a physically intuitive but rigorous representation of $\pi$-bands in the graphene layers. Within the Wannier framework, the band Hamiltonian is expressed in terms of the hopping energy strength and the Bloch function basis by $\ket{\psi_{\mathbf{k}\alpha}}=\frac{1} {\sqrt{N}}\sum_{R}e^{i\mathbf{k}\cdot(\mathbf{R+\tau_\alpha})} \ket{\mathbf{R+\tau_{\alpha}}}$, where $\alpha$ is the sublattice index, $\tau_{\alpha}$ is the position of the sublattice relative to the lattice vectors $\mathbf{R}$, and $\ket{\mathbf{R+\tau_{\alpha}}}$ is the Wannier function. The Hamiltonian matrix elements are related to the Wannier representation hopping amplitudes by $H_{\alpha\beta}(\mathbf{k})=\bra{\psi_{\mathbf{k}\alpha}}H\ket{\psi_{\mathbf{k}\beta}}$ and thereby $H_{\alpha\beta}(\mathbf{k})=\frac{1}{N}\sum_{RR'} e^{i\mathbf{k}\cdot(\mathbf{R}-\mathbf{R'})}t_{\alpha \beta}(\mathbf{R}-\mathbf{R'}) $, where $t_{\alpha \beta}(\mathbf{R}-\mathbf{R'})=\bra{\mathbf{R+\tau_{\alpha}}}H\ket{\mathbf{R'+\tau_{\beta}'}}$ represents hopping amplitudes tunneling from $\beta$ to $\alpha$ sub-lattice sites located at $\mathbf{R'+\tau_{\beta}'}$ and $\mathbf{R+\tau_{\alpha}}$ respectively. The single-particle Hamiltonian of bilayer $H(k)$ is in the form of
\begin{equation} \label{hami}
    H(\mathbf{k})=\begin{bmatrix} H^{l}(\mathbf{k}) & T(\mathbf{k})\\
T^\dag(\mathbf{k}) & H^{u}(\mathbf{k})
\end{bmatrix},
\end{equation}
where $H^{l(u)}$ is the Hamiltonian of the lower (upper) layer and $T(\mathbf{k})$ describes the interaction between the two layers. The Hamiltonian can be decoupled into contributions from intra-layer and inter-layer interactions, as well as intra-sublattice and inter-sublattice terms within each layer. The lattice vectors for each layer are $\mathbf{a}_{1}=a\left[\frac{1}{2},\frac{\sqrt{3}}{2}\right], \hspace{2mm}\mathbf{a}_{2}=a\left[-\frac{1}{2},\frac{\sqrt{3}}{2}\right]$ and the corresponding reciprocal lattice vectors $\mathbf{b}_{1}=\frac{2\pi}{\sqrt{3}a}\left[\sqrt{3},1\right],\hspace{2mm} \mathbf{b}_{2}=\frac{2\pi}{\sqrt{3}a}\left[-\sqrt{3},1\right]$, where the lattice constant $a=2.43${\AA}. We considered two stacking configurations of bilayer graphene (AA and AB) [Fig.\ref{fig:schem}]. To make the Haldane type \cite{PhysRevLett.61.2015}, we use the Haldane Hamiltonian, $H^h(\mathbf{k})=\mathbf{d^h(k)\cdot\sigma} + \epsilon^h \cdot \mathbb{I}$, where $\sigma$ is Pauli matrices, $d_x^h=t_g\sum_{i=1}^3 cos(\mathbf{k\cdot e_i})$, $d_y^h=-t_g\sum_{i=1}^3 sin(\mathbf{k\cdot e_i})$, $d_z^h=M-2t_2 sin(\phi)\sum_{i=1}^3 -(-1)^isin(\mathbf{k\cdot a_i})$ and $\epsilon_h=2t_2 cos(\phi)\sum_{i=1}^3 cos(\mathbf{k\cdot a_i})$. Here, M is the Semenoff mass, also known as the on-site potential; $t_{2}$ is the next nearest neighbor (NNN) hopping strength; $\phi$ is the NNN hopping phase; and $t_{g}$ is the nearest neighbor (NN) hopping strength. The vectors $\mathbf{e}_i$ and $\mathbf{a}_i$ link the nearest and next nearest neighbors to the honeycomb lattice, as shown in Fig.\ref{fig:schem}a. The Hamiltonian of the graphene layer without Haldane terms can be written using the simple relation $H^g(\mathbf{k})=\mathbf{d^g(k)\cdot\sigma}$, where $d^g_x=d^h_x$, $d^g_y=d^h_y$ and $d^g_z=0$. For monolayer graphene, the Haldane flux and staggered potential make the low-energy bands non-trivial \cite{PhysRevLett.61.2015} by breaking the time reversal and inversion symmetry. We explore the non-trivial nature of the bands forming a Graphene/Haldane ($\rm G/\tilde{G}$) bilayer, where the lower layer of the bilayer graphene is converted into a Haldane layer. The corresponding intra-layer Hamiltonians of both layers are converted to $H^l=H^h$, $H^u=H^g$\cite{PhysRevB.100.081107}. Furthermore, the Haldane/Haldane ($\rm \tilde{G}/\tilde{G}$) bilayer, where both layers are converted into Haldane layers with $H^l=H^h$, $H^u=H^h$. In our model, we considered only the vertical hopping interaction between the two layers to simplify the model, as illustrated in Fig.\ref{fig:schem}b. The coupling matrices between the two layers in the AA and AB stacking bilayer graphene were $T_{AA}=\gamma\mathbb{I}$ and $T_{AB}=\begin{bmatrix} 0 & 0 \\ \gamma & 0 \end{bmatrix}$ respectively, where $\gamma$ is the inter-layer hopping strength. We used the nearest-neighbor (NN) hopping strength, $t_g=-2.7$ eV (for both monolayer and bilayer graphene), and the inter-layer hopping strength, $\gamma=0.2t_g$ for both AA and AB stacking \cite{Wallace1947,Castro2009,Chittari_2019,PhysRevB.86.125413}. We use next nearest neighbor (NNN) hopping strength, $t_2=0.1t_g$ for monolayer and for bilayer either $t_2=0.1t_g$ or $t_2=0.01t_g$ with fixed $\phi = \pi/2$ value. We used the same values throughout the manuscript unless otherwise mentioned. Aside from graphene, other materials
described by the Haldane model have different lattice constants from graphene’s, one has to keep these concerns while constructing the
transfer matrix T is represented in Fig. 2(b). In the case of lattice mismatch,
graphene/non-graphene bilayer forms a moire superlattices~\cite{PhysRevLett.122.016401}. Further, we assumed that the lattice constants of the graphene and Haldane layers are same, so that the moire complexity is avoided
\subsection{Theory of Orbital Hall conductivity}\label{Sec:HC}
In crystals, inversion (\textit{I}) or time-reversal (\textit{TR}) symmetry breaking inherently induces non-vanishing Berry curvatures $\Omega_n(\mathbf{k})$, within the Brillouin zone. These Berry curvatures lead to an anomalous electron velocity in the presence of an external field \cite{PhysRevB.53.7010,PhysRevLett.95.137204}. Electrons with opposite velocities, orbital angular momentum, or spins generate a Hall voltage across opposite edges owing to the directional separation \cite{ARYASETIAWAN201987}. In the presence of broken \textit{TR} symmetry, uneven electron accumulation at the edges leads to the emergence of the anomalous Hall effect (AHE) \cite{RevModPhys.82.1959,PhysRevLett.95.137204,RevModPhys.82.1539}. By summing the Berry curvature $\Omega_{xy,n}^z(\mathbf{k})$, over the Brillouin zone (BZ), we can obtain the anomalous Hall conductivity as follows: 
\begin{equation}\label{AHC}
    \sigma_{xy}^z=\frac{1}{2\pi}\frac{e^2}{h}\sum_n \int d^2k \hspace{1mm}\Omega_{xy,n}^z(\mathbf{k}) f_{FD}^n
\end{equation}
where $f_{FD}^n=1/\left( e^{\frac{E_n-E_F}{K_BT}}+1\right)$ denotes the Fermi occupation function. Berry curvature of the $n_{th}$ energy band estimated from the formula,
\begin{equation}\label{Berry}
    \Omega_{xy,n}^z(\mathbf{k})=-2\sum_{n'\neq n}Im[\frac{\bra{\psi_n}\frac{\partial H}{\partial k_x} \ket{\psi_{n'}}\bra{\psi_{n'}}\frac{\partial H}{\partial k_y} \ket{\psi_n}}{(E_{n'}-E_n)^2}]
\end{equation}
Here, at each $\mathbf{k}$ point, we sum over all other bands ($n^{\prime}\neq n$). Where $\ket{\psi_n(\mathbf{k})}$ represents the Bloch states, and $E_n$ are their corresponding eigenvalues. The topological invariant bands are associated with non-vanishing Berry curvatures within the Brillouin zone (BZ) and are characterized by calculating the Chern number \cite{PhysRevLett.71.3697} of the $n_{th}$ band using $C_n=\frac{1}{2\pi}\int d^2(\Vec{k}) \Omega_n(\Vec{k})$.
\noindent
\par
Furthermore, in non-magnetic quantum materials,  an intrinsic orbital magnetic moment, denoted by $\mathbf{m}_n(\mathbf{k})$ for the n$^{th}$ band, emerges at equilibrium from the self rotation of the electron wave functions around their center of mass. According to the system Hamiltonian, the $\mathbf{m}_n(\mathbf{k})$ can be induced by either breaking the time-reversal (\textit{TR}) or inversion (\textit{I}) symmetry without any external stimuli \cite{PhysRevB.74.024408,PhysRevB.93.174417,PhysRevLett.110.087202}. Interestingly, the preserved symmetry was reflected in the orbital magnetic moment. For \textit{TR} symmetry, $\mathbf{m}_n(-\mathbf{k})=-\mathbf{m}_n(\mathbf{k})$ , and for \textit{I} symmetry, $\mathbf{m}_n(-\mathbf{k})=\mathbf{m}_n(\mathbf{k})$. 
In both cases, this leads to a net orbital magnetic moment of zero, that is, ($\mathbf{m}_n=0$). In non-magnetic quantum materials with broken symmetry, the induced orbital magnetic moment is complemented by the Berry curvature itself, which also exhibits the same behavior under the following symmetries: $\mathbf{\Omega}_n(-\mathbf{k})=-\mathbf{\Omega}_n(\mathbf{k})$ for \textit{TR} and $\mathbf{\Omega}_n(-\mathbf{k})=\mathbf{\Omega}_n(\mathbf{k})$ for \textit{I} \cite{PhysRevLett.99.236809}. In these systems, for an applied longitudinal electric field, oppositely oriented orbital magnetic moments accumulate at two opposite edges along the transverse direction, resulting in  the flow of orbital current. Correspondingly, one can obtain the orbital Hall conductivity, denoted $\sigma_{xy}^{orb}(E)$ in the following equation: 
\begin{equation} \label{orb_Hall}
    \sigma_{xy}^{z,orb}=-\frac{e}{(2\pi)^2}\sum_n \int d^2k   \hspace{1mm}\Omega_{xy,n}^{z,orb}(k) f_{FD}^n
\end{equation}
Here, $\Omega_{xy,n}^{z,orb}(k)$ is the orbital Berry curvature computed using the Kubo formula, 
\begin{equation}\label{OBC}
    \Omega_{xy,n}^{z,orb}(\mathbf{k})=2\hbar\sum_{n'\neq n}Im[\frac{\bra{\psi_n} \mathcal{J}_x^{z,orb} \ket{\psi_{n'}}\bra{\psi_{n'}} v_y \ket{\psi_n}}{(E_{n'}-E_n)^2}]
\end{equation}
and $\mathcal{J}_x^{z,orb}=\frac{1}{2}\{v_x,L^z\}$ represents the orbital current along the x-axis owing to the induced out-of-plane angular momentum ($L^z$) in the presence of an applied electric field along the y-axis. The induced orbital angular momentum (OAM) of the Bloch electron in a periodic system arises from the self-rotation of the wave packet, which comprises both intra- and inter-site contributions \cite{PhysRevLett.99.236809, PhysRevLett.95.137205,PhysRevB.108.075427}. From the Wannier representation, OAM is described by the Berry phase formula of the orbital magnetic moment as $L^z=-\frac{\hbar}{g_L\mu_B}m^z$ \cite{PhysRevLett.99.197202,PhysRevB.106.104414,PhysRevB.105.195421,PhysRevB.103.195309}, which simplifies the orbital current operator as follows:
\begin{equation}
    \resizebox{.9\hsize}{!}{$\bra{\psi_n(\mathbf{k})} \mathcal{J}_x^{z,orb} \ket{\psi_{n'}(\mathbf{k})} = -\frac{\hbar}{2g_L\mu_B}\bra{\psi_n(\mathbf{k})} v_xm^z+m^z v_x \ket{\psi_{n'}(\mathbf{k})}$} \nonumber
\end{equation} 
\begin{equation}
    \resizebox{.9\hsize}{!}{$= -\frac{\hbar}{2g_L\mu_B}[\sum_m \bra{\psi_n(\mathbf{k})} v_x \ket{\psi_{m}(\mathbf{k})} \bra{\psi_{m}(\mathbf{k})} m^z \ket{\psi_{n'}(\mathbf{k})} $} \nonumber
\end{equation}
\begin{equation} \label{OC}
    \resizebox{.7\hsize}{!}{$+ \bra{\psi_n(\mathbf{k})} m^z \ket{\psi_{m}(\mathbf{k})} \bra{\psi_{m}(\mathbf{k})} v_x \ket{\psi_{n'}(\mathbf{k})}]$}
\end{equation}
In the above Eq.(\ref{OC}), we have an intra-band ($m=n$) and inter-band ($m\neq n$) orbital magnetic moment ($m^z_{nm}$) contributions arises.\\ 
In general, the intrinsic orbital magnetic moment arises from the inter- and intra-band contributions of the Bloch bands of the system ~\cite{PhysRevB.105.195421}. However, the degenerate and non-degenerate Bloch bands can treated separately as discussed earlier~\cite{PhysRevLett.99.197202,PhysRevB.106.104414,PhysRevB.105.195421,PhysRevB.103.195309}. In the case of non-degenerate isolated $n^{th}$ Bloch band with a basis $\ket{\psi_n}$, the intrinsic orbital magnetic momentum can be obtained using a simple equation, Eq.(\ref{mz}), where $\bra{\psi_n}m^z\ket{\psi_m}=m_n^z\delta_{nm}$~\cite{PhysRevB.105.195421, PhysRevB.103.195309}. However, the nearly degenerate Bloch bands involve both the inter-and intra-bands, then the orbital magnetic moment should be calculated using the full equation, Eq.(\ref{OC})~\cite{PhysRevB.105.195421}. It was evident for the unbiased transition metal dichalcolgenides (TMDs)~\cite{PhysRevB.105.195421}, where the nearly degenerate conduction bands require the consideration of the non-Abelian structure of the orbital magnetic moments.
Here, we are dealing with the gapped graphene layers in the bilayer structure, where the gap is induced either by mass, $m$ term or NNN term, $t_2$, and the low-energy bands are well separated. The Hamiltonian produced the non-degenerate well separated Bloch bands, so, the inter-band matrix elements of the orbital magnetic moment, $\bra{\psi_n}m^z\ket{\psi_m}=0$ \cite{PhysRevB.103.195309} and only the (diagonal) intra-band components, ($m^z_n=m^z_{nn}$), contribute in the Eq.(\ref{OC}). Consequently, the orbital Berry curvature can be modified as follows:
\begin{equation}
  \resizebox{1\hsize}{!}{$
  \Omega_{xy,n}^{z,orb}(\mathbf{k})=-\frac{1}{g_L\mu_B}\sum_{n'\neq n}Im\left[(m_n+m_{n'})\frac{\bra{\psi_n} \frac{\partial H}{\partial k_x} \ket{\psi_{n'}}\bra{\psi_{n'}} \frac{\partial H}{\partial k_y} \ket{\psi_n}}{(E_{n'}-E_n)^2}\right]
  $}
\end{equation}
and the magnetic moment is given by
\begin{equation} \label{mz}
    m_n^z(\mathbf{k})=\frac{e}{2i \hbar}\bra{\nabla_k\psi_n(\mathbf{k})}\times[H(k)-E_n(K)]\ket{\nabla_k\psi_n(\mathbf{k})}
\end{equation}
By the unitary operation, $\mathbb{I}=\sum_{n'} \ket{\psi_{n'}(\mathbf{k})}\bra{\psi_{n'}(\mathbf{k})}$ we simplify the Eq.(\ref{mz}) as
\begin{equation}\label{mz_n}
    m_n^z(\mathbf{k})=-\frac{e}{\hbar} \sum_{n'\neq n} Im[\frac{\bra{\psi_n}\frac{\partial H}{\partial k_x} \ket{\psi_{n'}}\bra{\psi_{n'}}\frac{\partial H}{\partial k_y} \ket{\psi_n}}{E_{n'}-E_n}]
\end{equation}
The Eq.(\ref{mz_n}) suggests that the orbital magnetic moment arises from the Bloch states and is directly related to the band disperson, exhibiting an inverse proportionality to the energy difference. Because our tight-binding Hamiltonian assumes highly localized atomic orbitals, the orbital magnetic moment arises primarily from the dispersion of bands, rather than directly from the atomic orbitals themselves.
\begin{figure}
     \centering     
     \includegraphics[width=\columnwidth]{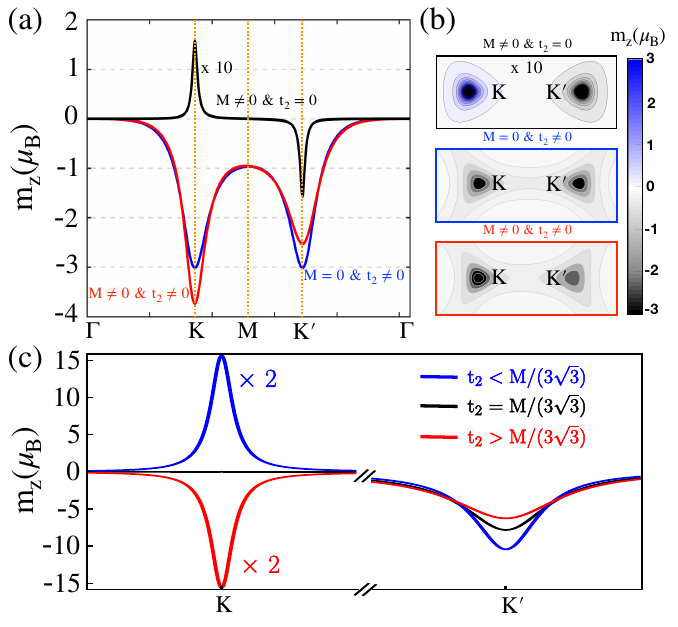}
     \caption{(a) The orbital magnetic moment ($m_z$) along the high symmetry path ($\Gamma \rightarrow K \rightarrow M \rightarrow K' \rightarrow \Gamma$) in the Brillouin zone (BZ). The black color line corresponding to the $m_z$ that arises with the inclusion of inversion symmetry breaking ($M=0.1t_g$) and no NNN hopping is added. The blue color line corresponding to $m_z$ that arises with the inclusion of time-reversal symmetry breaking with $t_2=0.1t_g$ and inversion symmetry preserved ($M=0$). For the case where both the symmetries are broken ($M=0.1t_g$ and $t_2=0.1t_g$) is represented with the red color line. Here we use $t_g=-2.7$ eV the nearest neighbor (NN) hopping strength and Haldane flux$(\phi)=\pi/2$. (b) The orbital magnetic moment distribution for all three cases around the BZ corners ($K$ and $K'$) is plotted. (c) The $m_z$ of single layer graphene for trivial ($t_2<M/3\sqrt{3}$), critical ($t_2=M/3\sqrt{3}$) and non-trivial ($t_2>M/3\sqrt{3}$) phases at $K$ and $K'$ valleys. }
    \label{fig:G_mag}
\end{figure}
\begin{figure*}
\centering
\includegraphics[width=\textwidth]{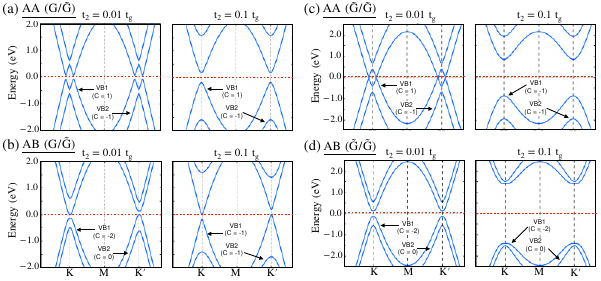}
\caption{Low-energy band structures of $\rm G/\tilde{G}$ and $\tilde{G}/\tilde{G}$ bilayers, where $M=0$, $\gamma=0.2t_g$, $\phi=\pi/2$ and $t_g=-2.7$ eV for both AA and AB stacking, respectively.}
\label{fig:G_G_bands}
\end{figure*}
\begin{figure*}
\centering
\includegraphics[width=\textwidth]{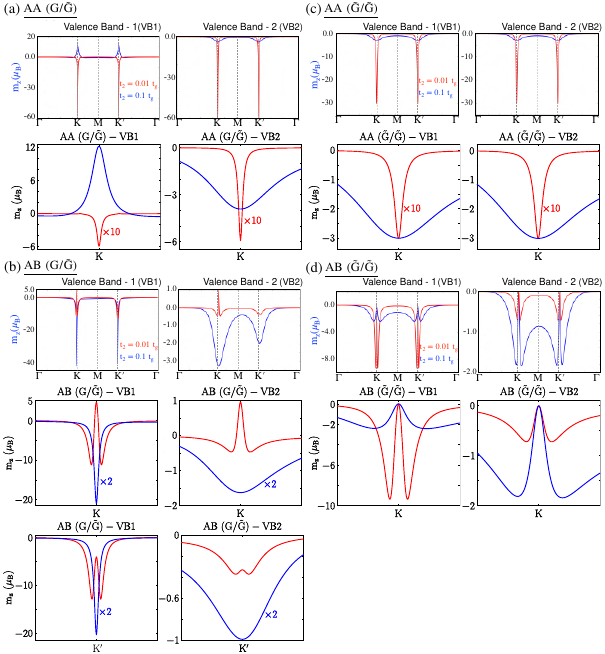}
\caption{(a,b): Orbital magnetic moment of the valence bands (Valence band-1, 2) of AA stacked and AB stacked $\rm G/\tilde{G}$ bilayers with highlighted $K-$valley distribution, where $M=0$, $\gamma=0.2t_g$, $\phi=\pi/2$ and $t_g=-2.7$ eV. The $m_z$ distribution at the $K'-$valley has identical as $K-$ valley. (c,d): $\rm \tilde{G}/\tilde{G}$ bilayer system, where $M_l=M_u=0$, $\gamma=0.2t_g$, $\phi_l=\phi_u=\pi/2$, $t_g=-2.7$ eV, and $t_{2l}=t_{2u}=t_2$ for two different NNN hopping strengths: $t_2=0.01t_g$(red line) and $t_2=0.1t_g$(blue line). The valley dependent $m_z$ of the $\rm AB(G/\tilde{G})$ and valley independent $m_z$ of $\rm AB(\tilde{G}/\tilde{G})$ are highlighted in each plots.}
\label{fig:G_G_mag}
\end{figure*}
 \begin{figure*}
     \centering
     \includegraphics[width=\textwidth]{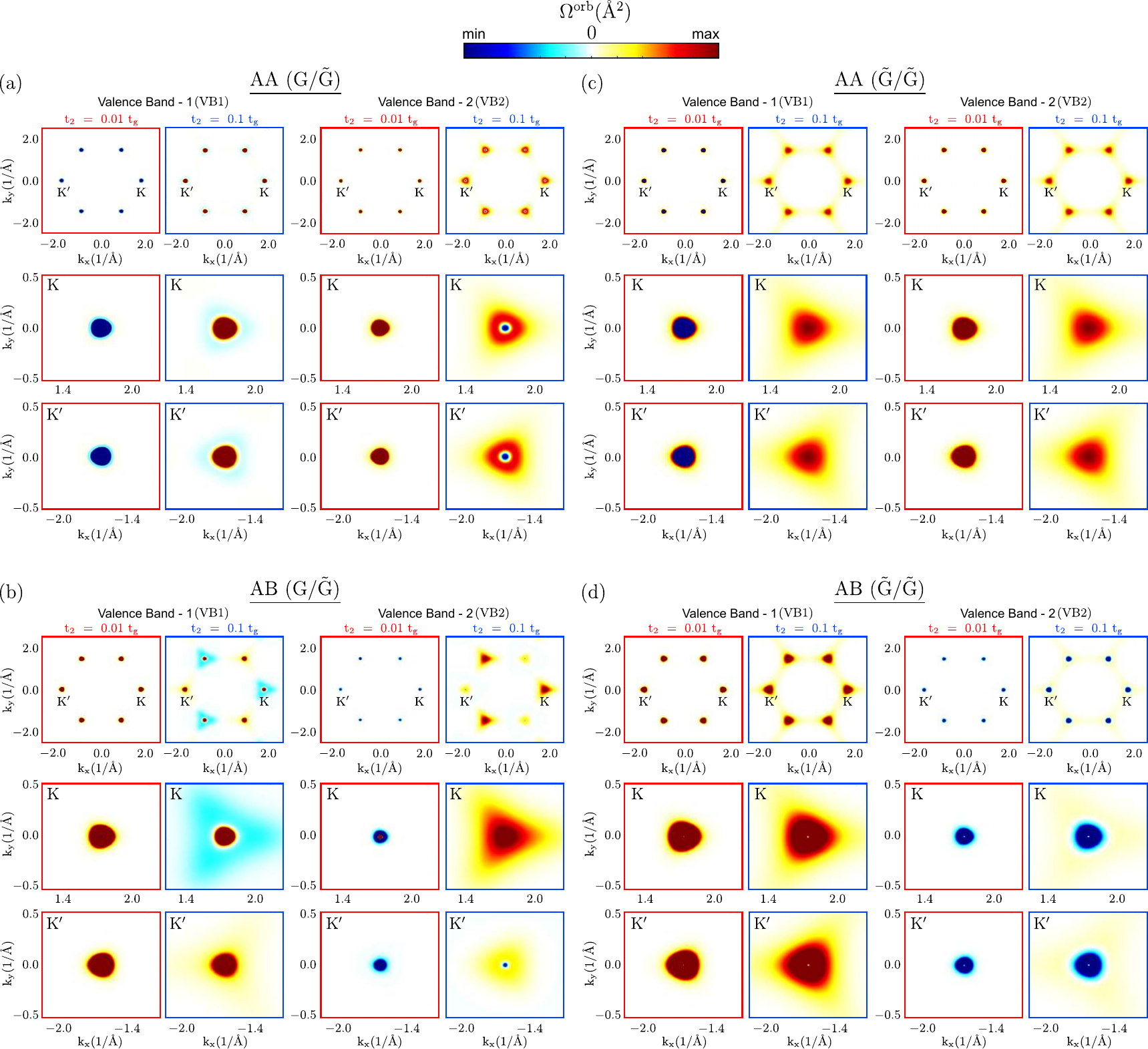}
     \caption{ (a,b) Orbital Berry curvatures of valence bands (Valence band-1,2) of AA stacked and AB stacked $\rm G/\tilde{G}$ bilayer, where $M=0$, $\gamma=0.2t_g$, $\phi=\pi/2$ and $t_g=-2.7$ eV and (c,d) $\rm \tilde{G}/\tilde{G}$ bilayer system, where $M_l=M_u=0$, $\gamma=0.2t_g$, $\phi_l=\phi_u=\pi/2$, $t_g=-2.7$ eV and $t_{2l}=t_{2u}=t_2$ for two different NNN hopping strength as $t_2=0.01t_g$ (red box) and $t_2=0.1t_g$ (blue box). For each case, the close distribution of orbital Berry curvature (OBC) around the K and $K'$ points is provided below.}
    \label{fig:OBC}
 \end{figure*}
\section{\label{sec:3}Results and Discussion}
\subsection{Orbital Ferromagnetism in Graphene}\label{Sec:SingleG}
We break the \textit{I} and \textit{TR} symmetries in single-layer graphene by introducing the mass term ($M$) and next nearest neighbor hopping term ($t_2$) (as discussed in Sec. \ref{TB_hamil}). This results in non-vanishing Berry curvatures in the bands. Consequently, the \textit{TR} or \textit{I} symmetry breaking induces the orbital magnetic moment, which is calculated using Eq.(\ref{mz_n}), which is consistent with previous studies \cite{PhysRevB.103.195309}. In Fig.\ref{fig:G_mag}(a), we present the calculated orbital magnetic moment carried by the Bloch states for breaking either one or both the \textit{I} and \textit{TR} symmetries. Fig.\ref{fig:G_mag}(b) shows the distribution of the orbital magnetic moment in the Brillouin zone (BZ) corners ($K$ and $K'$). We observed that the calculated $m^z$ was sensitive to the introduced symmetry breaking and strengths of $M$ and $t_2$. It should be noted that the mass term ($M=0.1t_g$) and the NNN term with $t_2 = 0$ open up a gap ($E_g=0.54$ eV) at the Dirac points ($K$ and $K'$) and breaks the \textit{I} symmetry. This results in a large orbital magnetic moment ($\sim$15$\mu_B$) induced, with opposite signs at $K$ and $K'$, as discussed previously in~Ref.\cite{PhysRevB.103.195309}. Remarkably, with the next nearest neighbor (NNN) term $t_2=0.1t_g$ ($M=0$), a large band gap ($E_g=2.81$ eV) is induced, which is five times larger than the value obtained with the mass term ($M\neq 0$) at Dirac points. Interestingly, the $m^z$ has the same polarity in both valleys ($K$ and $K'$) because of the broken \textit{TR} symmetry  (blue line in Fig.\ref{fig:G_mag}(a)). However, owing to the large gap, the induced orbital magnetic moment experienced a substantial reduction, reaching a value of -3$\mu_B$. Notably, we have a finite net orbital magnetic moment within the BZ and demonstrate {\it Orbital Ferromagnetism}. Additionally, the inclusion of both the (\textit{I} and \textit{TR}) broken symmetries ($M\neq 0,~t_2\neq0$) induces distinct band gap values at the Dirac points ($E_g= 2.27$ eV at $K$ and $E_g= 3.35$ eV at $K'$), consequently distinct orbital magnetic moments at Dirac points (-3.73$\mu_B$ at $K$ and -2.5$\mu_B$ at $K'$) yet with the same polarity observed in the graphene, as shown by the red line in Fig.\ref{fig:G_mag}(a). This is the valley-dependent orbital magnetic moment in single-layer graphene. Similar to the previous case, the net orbital magnetic moment within the BZ was still finite and exhibited {\it Orbital Ferromagnetism}. Notably, inclusion of the NNN term ($t_2$) is sufficient to observe orbital ferromagnetism in single-layer graphene. Furthermore, in the Haldane model the gap induced at the $K/K^{\prime}$ without a mass term is equals to $2(3\sqrt{3}t_2)$ \cite{PhysRevLett.61.2015}. It is relevant to discuss the evaluation of the $m_z$ using the gaps induced at $K/K^{\prime}$. It is known that, $M/t_2 = 3\sqrt{3}$ is a critical value for trivial to non-trivial transitions. Here, we consider $t_2$ values of $< M/3\sqrt{3}$ and $> M/3\sqrt{3}$, resulting in an equal gap ($E_g=0.27$ eV) at the $K$ for both trivial and non-trivial bands, respectively. For the trivial case, the $m_z$ polarization is opposite between the two valleys, as previously shown when $M\neq 0~\& ~t_2=0$. However, owing to the different energy gaps between the two valleys, $m_z$ had distinct values. In contrast, for the non-trivial case, $m_z$ exhibits the same polarity within the two valleys, similar to when $t_2\neq0$. We observed an identical $m_z$ value of $\sim 31 \mu_B$ with the same gap but opposite polarity at the $K$ in both trivial and non trivial cases, as illustrated in Fig.\ref{fig:G_mag}(c). Additionally, in the critical case where $t_2=M/3\sqrt{3}$, the bands cross at the $K$ point and the corresponding $m_z$ is zero. However, $m_z$ is $-7.8 \mu_B$ at the $K'$ with an energy gap of $E_g=1.08$ eV (see Fig.\ref{fig:G_mag}(c)). We list the magnetic moment values in the valleys along with the band gaps for each case in Table.\ref{table1} 
\begin{table}
 \caption{ \label{table1}The $m_z$ and the band gaps at each valley ($K/K^{\prime}$) for the three cases (trivial, critical and non-trivial).}
    \begin{tabular}{|c|c|c|c|c|}
        \hline
        \multirow{2}{*}{Haldane Phases} & \multicolumn{2}{|c|}{Gap~(eV)} & \multicolumn{2}{|c|}{$m_z$~($\mu_B$)}\\
        \cline{2-5}
        & $K$ & $K'$ & $K$ & $K'$ \\
        \hline
         $t_2<\frac{M}{3\sqrt{3}}$~(trivial)& 0.27 & 0.81 & 31.3 & -10.4 \\ 
         \hline
         $t_2=\frac{M}{3\sqrt{3}}$~(critical)& 0 & 1.08 & 0 & -7.8 \\
         \hline
         $t_2>\frac{M}{3\sqrt{3}}$~(non-trivial)& 0.27 & 1.35 & -31.3 & -6.26 \\
        \hline
    \end{tabular}
\end{table}
In the case of Haldane graphene bilayers, we only use the next nearest neighbor (NNN) term, $t_2$, and the mass term is zero ($M=0$) to observe the {\it Orbital Ferromagnetism}.
\subsection{Orbital Magnetism in Haldane graphene bilayers} 
For the Haldane graphene bilayers (denoted as $\rm G/\tilde{G}$ and $\rm \tilde{G}/\tilde{G}$), where one or both layers are of the Haldane type (represented by $\rm \tilde{G}$), we explored both the AA and AB stacking configurations for each case. In the case of a bilayer, we break the \textit{TR} symmetry by introducing a non-zero Haldane term ($t_2e^{i\phi}$) and setting the mass term ($M$) to zero in the Hamiltonian. As established in Section~\ref{Sec:SingleG}, the next nearest neighbor hopping term ($t_2$) is sufficient to induce orbital magnetic moments, thereby leading to the orbital Hall conductivity.
First, we extensively studied the energy spectra of the $\rm G/\tilde{G}$ and $\rm \tilde{G}/\tilde{G}$ bilayers for both the AA and AB stacking, as shown in Fig.\ref{fig:G_G_bands}.(a-d). 
\subsubsection{Bands and topology}\label{Sec:Bands}
In the AA stacked $\rm G/\tilde{G}$ bilayer, NNN $t_2\neq0$ in the lower layer, which is the Haldane type ($\rm \tilde{G}$), breaks the \textit{TR} symmetry in the bilayer and opens a gap at the Dirac points ($K$ and $K'$). For the NNN hopping strength $t_2=0.01t_g$ with $\phi=\pi/2$ in $\rm AA(G/\tilde{G})$, at two Dirac $K$ and $K'$ points, there is a band gap ($E_g = 0.14$ eV) opening, and all four bands, that is, two valence ($\rm VB1$, $\rm VB2$)  and two conduction ($\rm CB1$, $\rm CB2$) bands are isolated from each other (see Fig.\ref{fig:G_G_bands}.(a)). These isolated bands were found to be non-trivial and showed non-zero Chern numbers from the Berry curvatures, as shown in \ref{app.Berry}. The Chern numbers of valence bands $\rm VB1$ and $\rm VB2$ are $C = 1$  and $C = -1$, respectively. The Chern numbers of conduction bands $\rm CB1$ and $\rm CB2$ take the opposite signs of the valence band Chern numbers as $C = -1$  and $C = 1$, respectively. For increased values of the $t_2$, that is, $t_2=0.1t_g$, there is an enhanced band gap ($E_g = 0.37$ eV), and all the bands are well separated from each other with a quadratic nature near the Dirac points, as shown in Fig.\ref{fig:G_G_bands}.(a). However, the Chern numbers of $\rm AA(G/\tilde{G})$ remained unaltered with $t_2=0.1t_g$. Interestingly, with the stacking transformation from AA to AB-, in the $\rm G/\tilde{G}$ bilayer, the valence band maximum touches the Fermi level at the Dirac point, $K$, and the conduction band minimum touches the Fermi level at the Dirac point, $K'$ (see Fig.\ref{fig:G_G_bands}.(b)). The asymmetric band gap opening at the Dirac points in $\rm AB(G/\tilde{G})$ bilayer creates valley-dependent band gaps with a p-type semiconducting nature at $K$ and n-type semiconducting nature at $K'$. Moreover, the bands in $\rm AB(G/\tilde{G})$ bilayer are separated:  yet we see that only two bands ($\rm VB1$ and $\rm CB1$) near the charge neutrality are non-trivial with Chern numbers $C = -2$ (for $\rm VB1$) for $t_2=0.01t_g, \phi=\pi/2$. For $t_2=0.1t_g$, the valley-dependent band gap opening is still present, and the Chern number of $\rm VB1 (CB1)$ changes $C = -2 (2)\rightarrow -1 (1)$. Further, the trivial valance band ($\rm VB2$) becomes non-trivial with Chern number $C =-1$. It is to be noted that, in the above two (AA and AB) stacking configurations of $\rm G/\tilde{G}$, the higher energy bands $\rm VB2$ and $\rm CB2$ are pushed to further higher energy with the increased value of $t_2$.  \par We now discuss the case in which the upper and lower layers are Haldane type ($\rm \tilde{G}$). Here, we take the same $t_2$ and $\phi$ values for both layers as $t_{2l}=t_{2u}=t_2$ and $\phi_l=\phi_u=\pi/2$, and the energy spectra for both AA- and AB- stacked $\rm \tilde{G}/\tilde{G}$ are shown in Fig.\ref{fig:G_G_bands}.(c,d). The upper and lower layers have different values of the Haldane terms $t_2$ and $\phi$ in  the bilayers. Their key role in the band topology of the $\rm \tilde{G}/\tilde{G}$ bilayer is discussed in \ref{app:ChernPhase}. For the AA-stacked $\rm \tilde{G}/\tilde{G}$ bilayer, the bands are nearly degenerate near the Dirac points ($K$ and $K'$) with a narrow band gap ($E_g = 0.66$ meV), as shown in Fig.\ref{fig:G_G_bands}.(c). Similar to AA($\rm {G}/\tilde{G}$) bilayer, in this case, all four bands are also topological for $t_2=0.01t_g$ with $C = 1 (-1)$ for $\rm VB1(VB2)$. Further, these bands become quadratic with increasing value of $t_2=0.1t_g$ (see Fig.\ref{fig:G_G_bands}.(d)). Therefore, the Chern numbers have been modified to $C = 1\rightarrow -1$ for $\rm VB1$ and $C = -1$ remains the same for $\rm VB2$. The conduction band Chern numbers are the same as those of the valence bands with opposite signs. In the case of the AB-stacked $\rm \tilde{G}/\tilde{G}$ bilayer, the bands are quadratic and have a clear band gap ($E_g =0.28$ eV) at the charge neutrality point, which increases with the $t_2$. However, the AB stacked $\rm \tilde{G}/\tilde{G}$ bilayer has only one topological band ($\rm VB1$) with Chern number $C = -2$ and remains unaltered with the increase in NNN hopping strength $t_2=0.01t_g \rightarrow 0.1t_g$. For the AA and AB stacking configuration of $\rm \tilde{G}/\tilde{G}$ with the increasing value of $t_2$, both the valence and conduction bands were pushed to higher energy, unlike the case of $\rm {G}/\tilde{G}$.  As discussed earlier, the presence of a single Haldane layer in $\rm G/\tilde{G}$ breaks \textit{TR} symmetry across the entire bilayer and leading to non-trivial low-energy bands. Similarly, when both layers in the bilayer were Haldane layers ($\rm \tilde{G}/\tilde{G}$), the bands became enriched with topological characteristics. Interestingly, a topological phase transition is also observed in both $\rm G/\tilde{G}$ and $\rm \tilde{G}/\tilde{G}$ with increasing $t_2$. However, the critical value of $t_2$ for the transition depended on the stacking type (AB or AA).
\subsubsection{Orbital magnetic moment}\label{Sec:Bi-mag}
A non-zero orbital magnetic moment arises with \textit{TR} or \textit{I} symmetry breaking. The topological bands in $\rm G/\tilde{G}$ and $\rm \tilde{G}/\tilde{G}$ due to \textit{TR} symmetry breaking are expected to show a non-zero orbital magnetic moment. We calculated the orbital magnetic moment distribution within the BZ for the AA and AB stacked $\rm G/\tilde{G}$ and $\rm \tilde{G}/\tilde{G}$ bilayers, as shown in Fig.\ref{fig:G_G_mag}.(a-d). The corresponding orbital Berry curvature distributions in the BZ are presented in Fig.\ref{fig:OBC}(a-d). The orbital Berry curvature hot spots are narrow at the Dirac points ($K$ and $K'$); thus, the orbital magnetic moment distribution is also mostly situated around the Dirac points. For $\rm VB1$ of $\rm AA (G/\tilde{G})$ bilayer, the orbital magnetic moment ($m_z$) is found to be -60$\mu_B$ at Dirac points ($K$ and $K'$) with $t_2=0.01t_g$, and it is reduced to ~12$\mu_B$ and changes its sign with increasing value of $t_2=0.1t_g$ (see Fig.\ref{fig:G_G_mag}.(a)). The orbital magnetic moment for $VB1$ has the same polarity at the two Dirac points ($K$ and $K'$) for $\rm AA (G/\tilde{G})$ bilayer, which is consistent with the increasing value of $t_2$ with the opposite polarity. It is to be noted that, for the $\rm VB2$ of $\rm AA (G/\tilde{G})$ bilayer, the orbital magnetic moment retains the same polarity for both $t_2=0.01t_g$ and $t_2=0.1t_g$ at two Dirac points. The orbital magnetic moment was -60$\mu_B$ with $t_2=0.01t_g$. same as $\rm VB1$ and is reduced to a small value of ~-4$\mu_B$ with $t_2=0.1t_g$, as shown in Fig.\ref{fig:G_G_mag}.(a). However, for $\rm AB (G/\tilde{G})$ bilayer show smaller value of orbital magnetic moment $5(1)\mu_B$ and $-4(-0.34)\mu_B$ at $K$ and $K'$, with $t_2=0.01t_g$ for $\rm VB1(VB2)$, respectively. Interestingly, unlike the $\rm AA (G/G)$ bilayer, the orbital magnetic moment polarity has opposite signs with different strengths at the two Dirac points, $K$ and $K'$, unlike the $\rm AA (G/\tilde{G})$ bilayer. This can be attributed to the interplay between the next nearest neighbor hopping of the lower layer and the AB stacking configuration. In the AB-stacked $\rm G/\tilde{G}$ bilayer, next nearest neighbor hopping of the lower layer perturbs only one sublattice of the upper graphene layer, which differs in the magnitude and polarization of the magnetic moments induced at the valleys, leading to a {\it valley-dependent orbital magnetic moment.} Moreover, the orbital magnetic moment distribution of $\rm VB1(VB2)$ shows oscillatory behavior with positive and negative values around the $K$ point and has a negative value of $m_z$ at $K'$ for $t_2=0.01t_g$ (see Fig.\ref{fig:G_G_mag}.(b)).This is evident from the orbital Berry curvatures, as shown in Fig.\ref{fig:OBC}(b). It should be noted that with increasing value of $t_2 = 0.1t_g$, the $m_z$ is enhanced to $-42(-3.25)\mu_B$ and $-41(-2)\mu_B$ at $K$ and $K'$ for $\rm VB1(VB2)$. For both layers, the Haldane type and, $\rm \tilde{G}/\tilde{G}$ bilayer with AA and AB stacked configurations showed no valley dependency on the $m_z$. An identical orbital magnetic moment ($-30\mu_B$ at Dirac points) behavior was observed for $\rm VB1$ and $\rm VB2$ in the $\rm AA\tilde{G}/\tilde{G}$ bilayer and reduced to smaller values ($-2\mu_B$) with increasing value of $t_2$ (see Fig.\ref{fig:G_G_mag}.(c)). The orbital Berry curvature hot spots change their sign with increasing $t_2$ for $\rm VB1$, which is not the case for $\rm VB2$, as shown in Fig.\ref{fig:OBC}(c). In the $\rm AB\tilde{G}/\tilde{G}$ bilayer, the magnitude of $m_z$ decreased for $\rm VB1$ with an increased value of $t_2$, whereas for $\rm VB2$ it increased, as shown in Fig.\ref{fig:G_G_mag}.(d).
\begin{figure}
     \centering     
     \includegraphics[width=0.8\columnwidth]{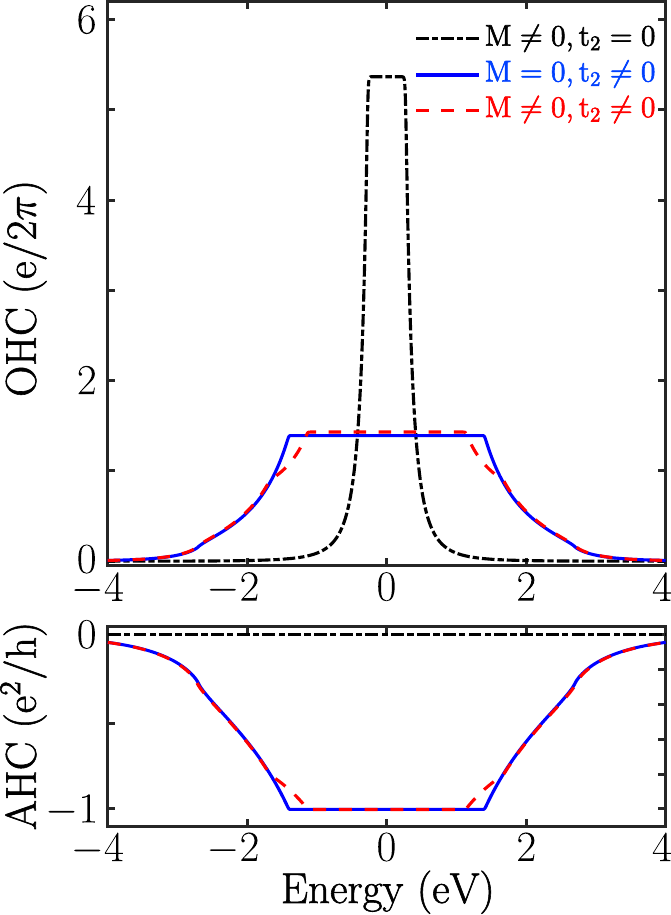}
     \caption{Orbital Hall conductivity (OHC) and anomalous Hall conductivity (AHC) for the three symmetry breaking cases, where top one represent the OHC in unit of $e/2\pi$ and bottom one represent the AHC in units of $e^2/h$. The black color line corresponding only inversion symmetry breaking ($M=0.1t_g$), blue line and red line corresponding time-reversal symmetry breaking with $t_2=0.1t_g$ and both the symmetry breaking with $M=0.1t_g$, $t_2=0.1t_g$ respectively. 
     }
    \label{fig:G_hall}
\end{figure}
\subsection{Orbital Hall conductivity}
\subsubsection{Single layer graphene}
It is interesting to observe the behavior of orbital Hall conductivity (OHC) in single-layer graphene. As discussed in Section.\ref{Sec:HC}, under an applied in-plane electric field, the oppositely oriented orbital magnetic moment ($m_z$) that accumulates at opposite edges yields the OHC. The OHCs calculated using Eq.(\ref{orb_Hall}) is shown in Fig.\ref{fig:G_hall} and is compared with the anomalous Hall conductivity (AHC). For cases $M\neq0~\&~t_2=0$, the AHC does not contribute. However, a very large OHC quantization was observed within the gap. Because no \textit{TR} symmetry is broken ($t_2=0$), the bands remain trivial and the opposite orbital moments at the valleys ($K$ and $K'$) contribute to the orbital current. This phenomenon is ascribed to the valley orbital Hall effect \cite{PhysRevB.103.195309}. In the case where the \textit{TR} symmetry is broken ($t_2\neq0$) i.e., for $M=0~\&~t_2\neq0$ and $M=0~\&~t_2\neq0$ and $M\neq0~\&~t_2\neq0$, the AHC is quantized to the value of $-e^2/h$ in the gap, and the bands are topological. Further, the OHC is also found to be quantized in the gap. Here, the bands are topological and show quantized AHC and OHC, which can be attributed to an {\it Orbital Chern Insulator}.
\begin{figure}[h!]
    \centering   
          \includegraphics[width=\columnwidth]{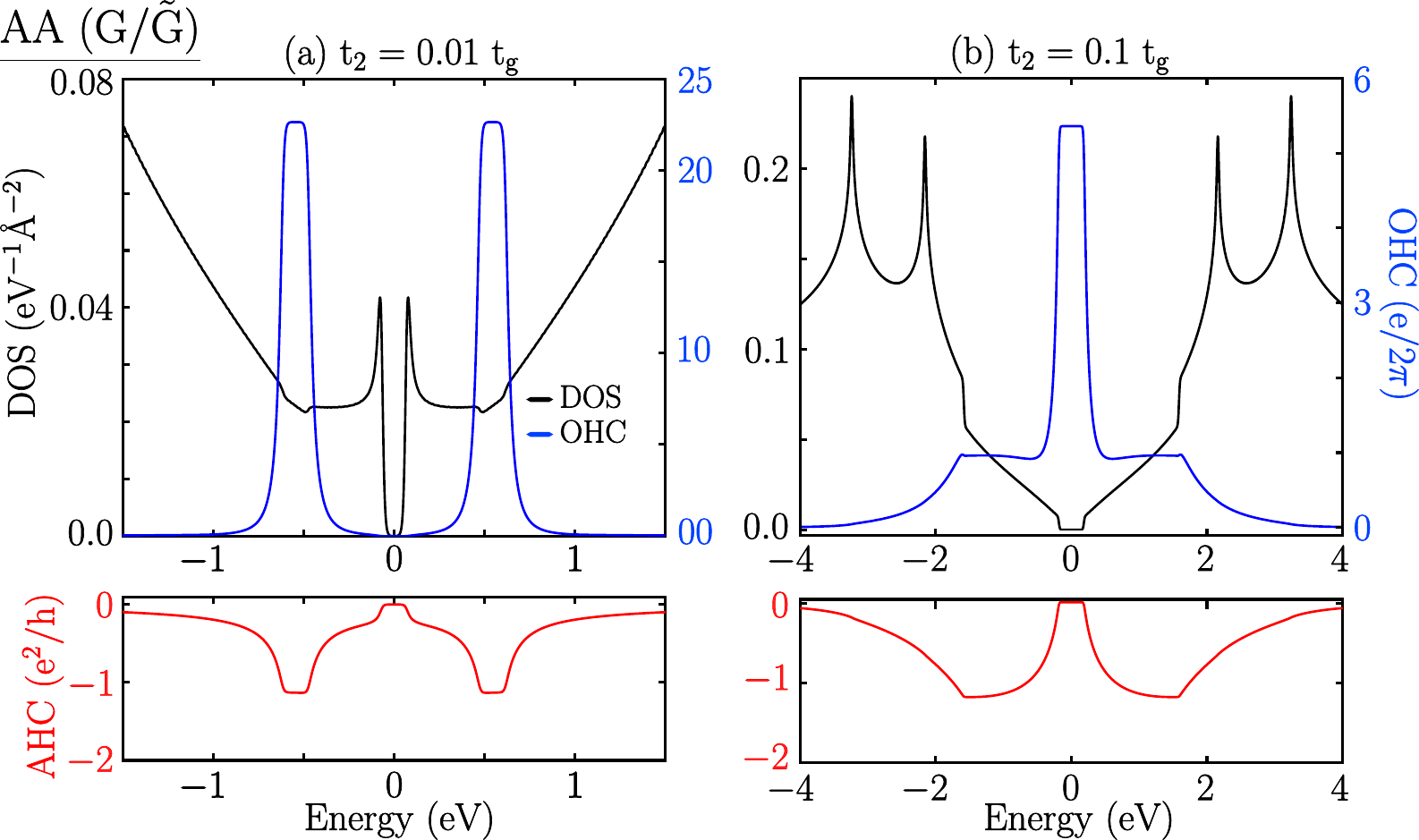}
        \includegraphics[width=\columnwidth]{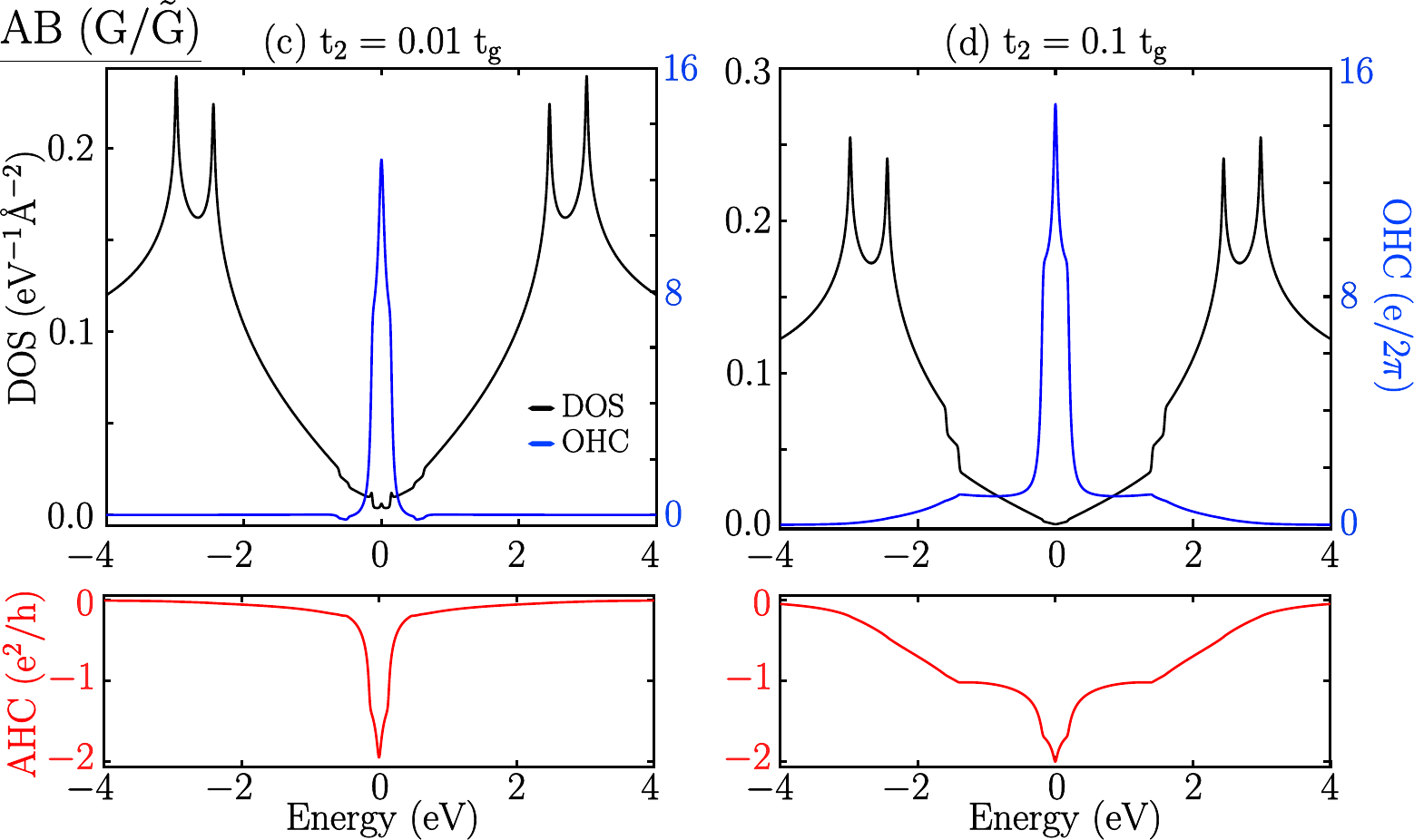}
    \caption{The density of states (DOS) and orbital Hall conductivity (OHC)(in units of $e/2\pi$) as a function of the Fermi energy for AA stacked (a,b) and AB stacked(c,d) of $\rm G/\tilde{G}$ bilayers. We also presented the anomalous Hall conductivity (AHC)(in units of $e^2/h$) for the comparison in red color in the lower panel. The OHC is sensitive to the NNN hopping strength. The first column is for $t_2=0.01t_g$ (a,c), and the second column corresponds to $t_2=0.1t_g$ (b,d).}
    \label{fig:G_H_hall}
\end{figure}
\subsubsection{$\rm G/\tilde{G}$}
The $\rm G/\tilde{G}$ and $\rm \tilde{G}/\tilde{G}$ bilayers showed different orbital magnetic polarization and valley-dependent orbital magnetic moments at the BZ corners ($K$ and $K'$) for both AA and AB stacking configurations; thus, the OHC has revealed various quantum phases of the Haldane graphene bilayers. We calculated the OHC for $\rm G/\tilde{G}$ and $\rm \tilde{G}/\tilde{G}$ bilayers and compared them with the AHC along with the density of states, as shown in Fig.\ref{fig:G_H_hall} and \ref{fig:H_H_hall}. The OHC of $\rm AA (G/\tilde{G})$ bilayer has quantized values wherever the AHC is quantized for $t_2 = 0.01t_g$ (see Fig.\ref{fig:G_H_hall}(a)). The quantized value of the AHC is slightly higher than unity because of the contribution of the partially filled bands along with the fully filled bands at a given Fermi energy. Both are quantized at the gap between the bands. Moreover, in the gap at charge neutrality, neither OHC nor AHC is quantized and becomes zero. As discussed earlier, AHC follows the accumulation of the occupied individual band's Chern number ($\rm VB1$ and $\rm VB2$ are 1 and -1, respectively) and shows a zero value at the charge neutrality point. Moreover, the orbital magnetic moments of the $\rm VB1$ and $\rm VB2$ bands have the same polarity, and the accumulation leads to zero OHC at the charge neutrality point. Interestingly, for $t_2=0.1t_g$, AHC has a zero value for the energy gap at charge neutrality. However, OHC has a quantization value, indicating that a new phase can be called a {\it Quantum Orbital Hall Insulator}. For $t_2=0.1t_g$, the Chern numbers of $\rm VB1$ and $\rm VB2$ remain the same as for $t_2=0.01t_g$, and the AHC has zero value at charge neutrality. It should be noted that, unlike the case of $t_2=0.01t_g$, the induced orbital magnetic moments of the individual bands have opposite polarities. This can be attributed to the transformation of the VB1 band from a `Mexican hat' behavior to a quadratic behavior as $t_2$ increases. This change in band character caused the orbital magnetic moment polarization for the VB1 band at the BZ corners to flip direction and accumulate, leading to quantized OHC, as shown in Fig.\ref{fig:G_H_hall}(b). In both cases ($t_2=0.01t_g$ or $t_2=0.1t_g$), the OHC is quantized in the gap between bands whenever they filled. For the AB stacked $\rm G/\tilde{G}$ bilayer, the OHC and AHC peaks at the charge neutrality point as the overall gap is zero (there is a gap at each Dirac point) for both the values of $t_2=0.01t_g$ and $t_2=0.1t_g$ as illustrated in Fig.\ref{fig:G_H_hall}(c) and (d). 
\subsubsection{$\rm \tilde{G}/\tilde{G}$}
\begin{figure}[h]
    \centering   \includegraphics[width=\columnwidth]{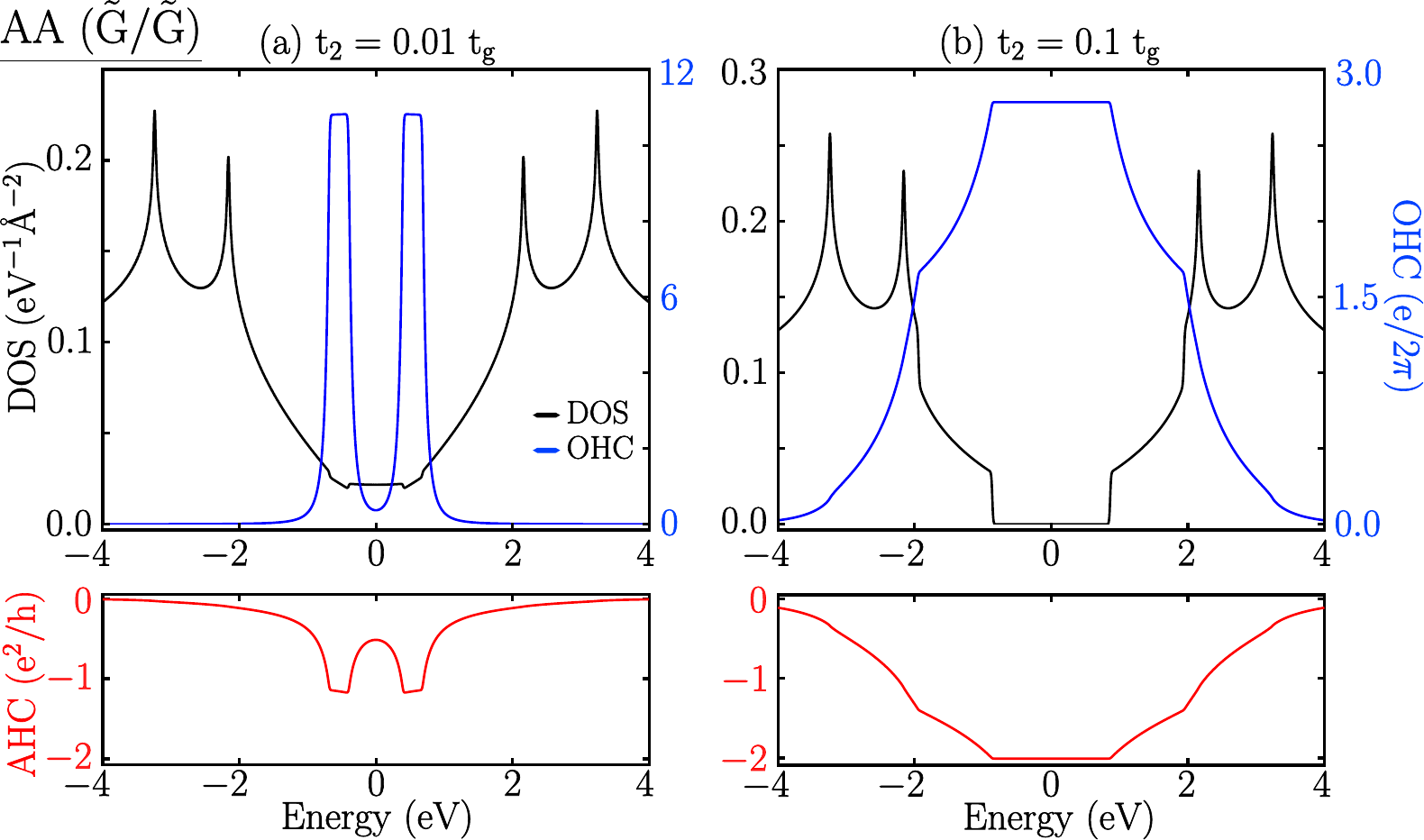}
    \includegraphics[width=\columnwidth]{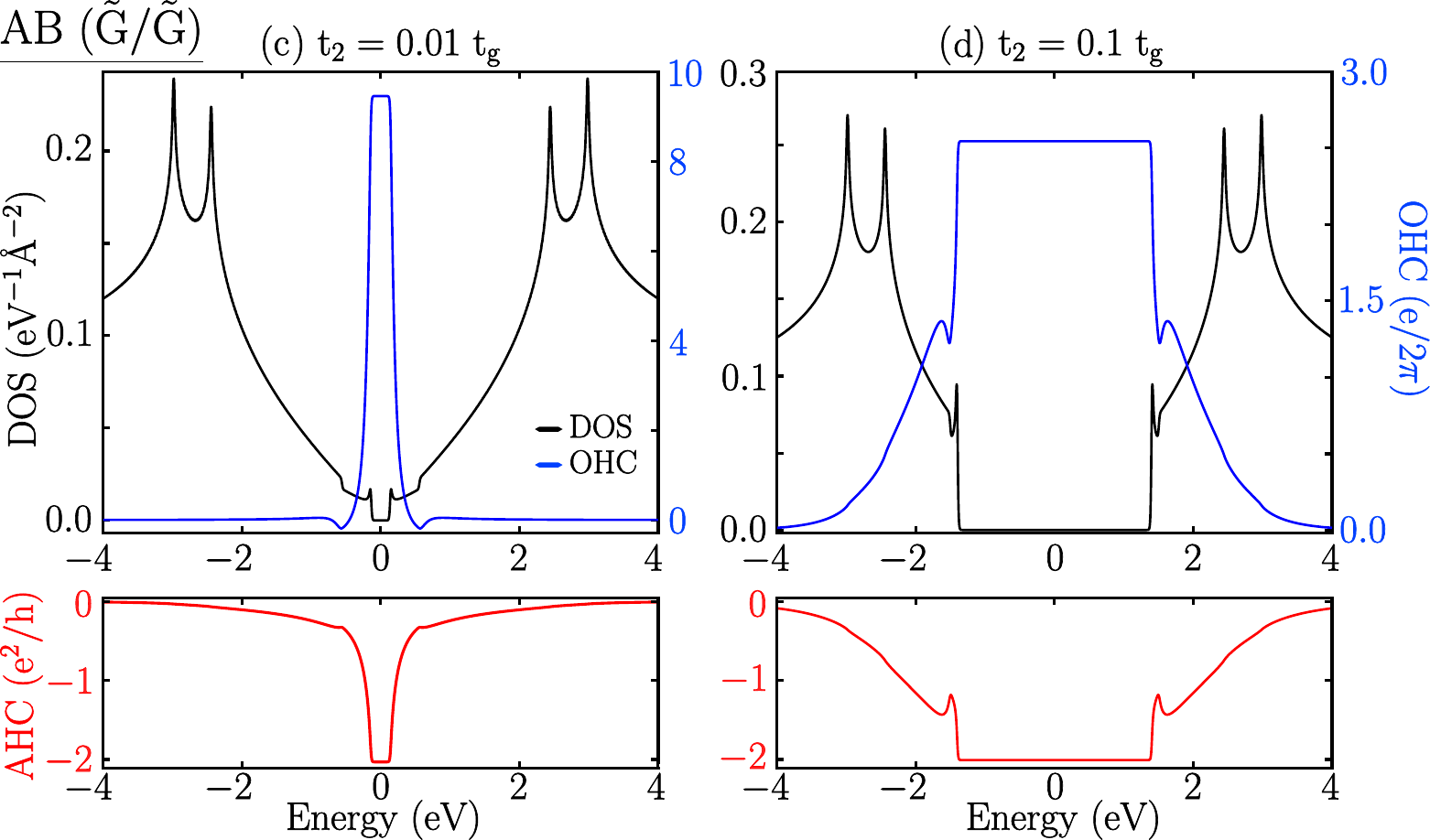}
    \caption{The density of states (DOS) and orbital Hall conductivity (OHC)(in units of $e/2\pi$) as a function of the Fermi energy for AA stacked (a,b) and AB stacked(c,d) of $\rm \tilde{G}/\tilde{G}$ bilayers. The lower panel with a red line represents the anomalous Hall conductivity (AHC)(in units of $e^2/h$). The first column is for $t_2=0.01t_g$ (a,c), and the second column corresponds to $t_2=0.1t_g$ (b,d).}
    \label{fig:H_H_hall}
\end{figure}

For $\rm \tilde{G}/\tilde{G}$ bilayers, the $\rm AA (\tilde{G}/\tilde{G})$ bilayers with $t_2 = 0.01t_g$ show similar behavior to OHC and AHC, as shown in Fig.\ref{fig:H_H_hall}(a). Interestingly, for $t_2 = 0.1t_g$, the $\rm AA (\tilde{G}/\tilde{G})$ shows quantized OHC and AHC in the gaps at charge neutrality (see Fig.\ref{fig:H_H_hall}(b)). Similarly, $\rm AB (\tilde{G}/\tilde{G})$ bilayers for both the values of $t_2=0.01t_g$ and $t_2=0.1t_g$ show quantized OHC and AHC in the gaps at charge neutrality (see Fig.\ref{fig:H_H_hall}(b,d)). In the non-gap region, OHC always follows the AHC trend. Both OHC and AHC quantization upon filling the occupied states lead to an {\it Orbital Chern Insulator}.
Quantum Orbital Hall phases arise in non-magnetic quantum materials with \textit{TR} breaking, as demonstrated in Haldane graphene bilayers. The induced orbital magnetic moment from the bands gives rise to the emergence of Orbital Magnetism and Valley Orbital Magnetic moment, consequently resulting in the manifestation of the Quantum Orbital Hall Effect and Orbital Chern insulator behavior. One can also demonstrate these Orbital Quantum Hall phases in the hetero structures. In \ref{app.Hetero}, we present the results of the Triangular/$\tilde{G}$, Kagome/$\tilde{G}$ heterostructures with one of the layers being Haldane type, generating the orbital Hall conductivity.
\section{\label{sec:4}Conclusion}
We demonstrated that the isolated bands in Haldane graphene bilayers, due to \textit{I} or \textit{TR} symmetry breaking, are of an orbital nature and have various Quantum Orbital Hall (QOH) phases. Although AHC and OHC qualitatively behave similarly in the Haldane single layer with \textit{TR} symmetry breaking, the scenario is different for the bilayers. The interplay of the NNN hopping strength and stacking configuration induces the Orbital Hall Insulator phase, where the OHC and AHC behave differently. Importantly, the Orbital Chern Insulator phase was always present in both layers. Furthermore, we demonstrated that valley dependent orbital magnetism arises by tuning the NNN hopping strength in the AB stacked Haldane graphene bilayers. Orbital magnetism in 2D layers is an emergent phenomenon, and our findings provides an understanding of its origins in simple Haldane graphene bilayers. 
\section{Acknowledgments}
We acknowledge the support provided by the Kepler Computing Facility, maintained by the Department of Physical Sciences, IISER Kolkata, for various computational requirements. S.G. acknowledges the support from the Council of Scientific and Industrial Research (CSIR), India, for the doctoral fellowship. B. L. C acknowledges the SERB with Grant No. SRG/2022/001102 and ``IISER Kolkata Start-up-Grant" No. IISER-K/DoRD/SUG/BC/2021-22/376.
%
\appendix
\section{Berry curvatures}\label{app.Berry}
\begin{figure}
    \centering
    \includegraphics[width=\columnwidth]{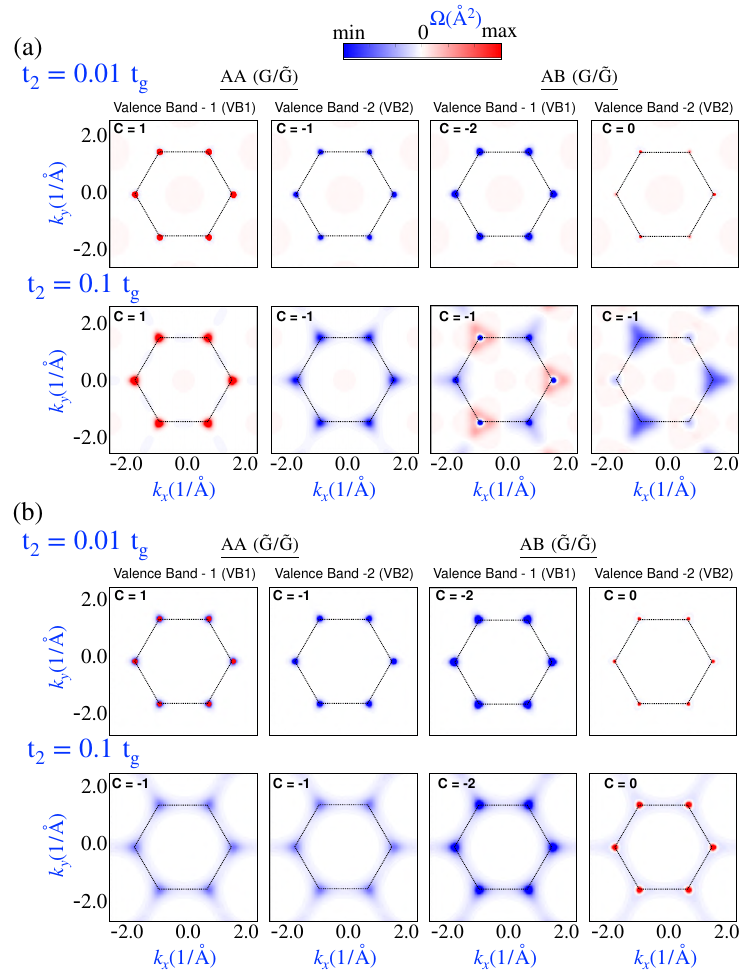}
    \caption{Berry curvature of valence bands (Valence band-1,2) of AA stacked and AB stacked (a) $\rm G/\tilde{G}$ and (b) $\rm \tilde{G}/\tilde{G}$ bilayer system for two different NNN hopping strength as $t_2=0.01t_g$ and $t_2=0.1t_g$. The typical values in the $\rm G/\tilde{G}$ bilayer of the chosen parameters are $M=0$, $\gamma=0.2t_g$,$\phi=\pi/2$ and $t_g=-2.7$ eV and for the $\rm \tilde{G}/\tilde{G}$ bilayer, the parameters are $M_l=M_u=0$, $\gamma=0.2t_g$, $\phi_l=\phi_u=\pi/2$, $t_g=-2.7$ eV and $t_{2l}=t_{2u}=t_2$.}
    \label{App:BerryCurv}
\end{figure}

The Berry curvatures of all the isolated bands for the $\rm G/\tilde{G}$ and $\rm \tilde{G}/\tilde{G}$ bilayers with the AA and AB stacking configurations with $t_2=0.01t_g$ and $t_2=0.1t_g$ are shown in Fig.\ref{App:BerryCurv}. Owing to the narrow band gap opening at the Dirac points ($K$ and $K'$) with the next nearest neighbor (NNN) hopping term $t_2e^{\pm i\phi}$, the Berry curvature hot spots are mainly localized at the BZ corners. All the hot spots at $K$ and $K'$ points are asymptotic with large positive or negative values, except in the case of AB stacked $\rm G/\tilde{G}$ where we have the p-type and n-type semiconducting nature of $K$ and $K'$ points. Moreover, the signatures of the Berry curvatures resemble those of the orbital magnetic moment and orbital Berry curvatures as shown in Fig.\ref{fig:G_G_mag} and \ref{fig:OBC}, respectively. 
\section{Chern phase diagrams}\label{app:ChernPhase}
\begin{figure}[h]
     \centering
     \includegraphics[width=\columnwidth]{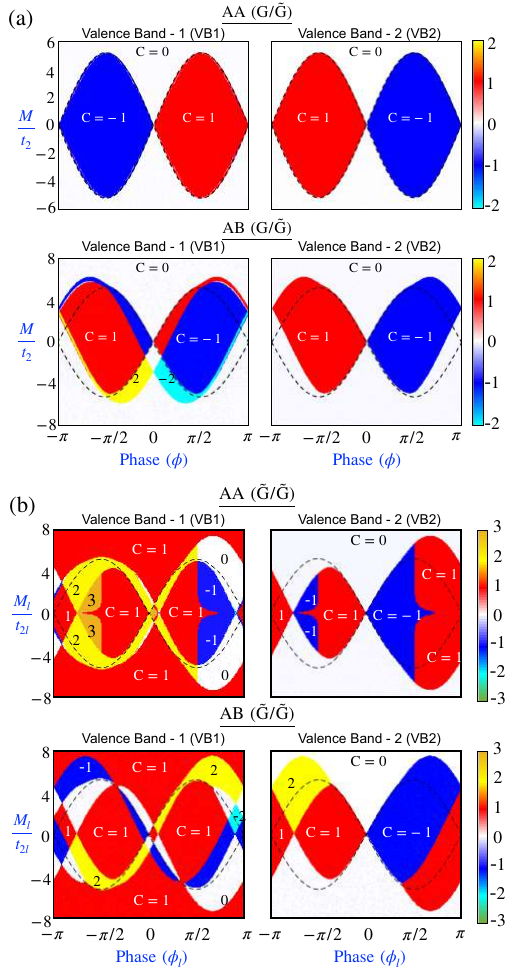}
     \caption{Chern phase diagram of the valance bands of AA stacked and AB stacked (a) $\rm G/\tilde{G}$ and (b) $\rm \tilde{G}/\tilde{G}$ bilayer, plotted as a function of the ratio of the lower layer onsight potential energy $M$ to the lower layer NNN hopping term $t_2$ against the Haldane flux $\phi$. The (dashed) black loop corresponding to the single Haldane layer. In the $\rm \tilde{G}/\tilde{G}$ bilayer, we fixed the parameters values of upper layer as $M_u=0, t_{2u}=0.1t_g$ and $\phi_u=\pi/2$. }
    \label{fig:ChernPhase}
 \end{figure}

\begin{figure}
    \centering
    \includegraphics[width=\columnwidth]{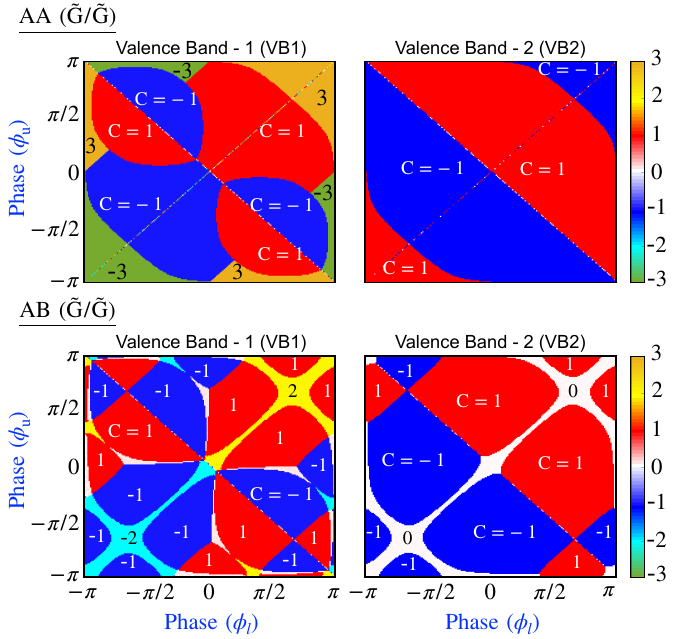}
    \caption{Chern phase diagram of the valance bands of AA stacked and AB stacked  $\rm \tilde{G}/\tilde{G}$ bilayer, plotted as a function of the lower Haldane flux $\phi_l$ to upper Haldane flux $\phi_u$, where $t_{2l}=t_{2u}=0.1t_g$, and $M_l=M_u=0$. }
    \label{fig:chern_phi_l_phi_u}
\end{figure}

The Chern phase diagram of the valance bands of AA and AB stacked of $\rm G/\tilde{G}$ bilayers and comparison with the Haldane type single layer are shown in Fig.\ref{fig:ChernPhase}(a). We present the Chern number as a function of the ratio of the on-site potential energy $M$ to the NNN hopping term $t_{2}$ against the Haldane flux $\phi$. For $\rm AA (G/\tilde{G})$ bilayer, the Chern phase diagram of $\rm VB1$ and $\rm VB2$ is shown with a colored surface, and the black dashed line represents a single Haldane layer (see Fig.\ref{fig:ChernPhase} (a)). Topological phase boundary of $\rm AA(G/\tilde{G})$ bilayer exactly matches the single Haldane layer Chern phase diagram. The chern number value $C=\pm 1$ was the same as that of the single Haldane layer. The $\rm VB2$ takes same values as the $\rm VB1$ Chern number values with opposite signs. We  observed deviations from the Haldane single layer Chern phase diagram for the $\rm AB (G/\tilde{G})$ bilayer, indicating the geometrical dependence of the bands topology in bilayer graphene. The Chern numbers of $\rm VB1$ vary between $C = 0, \pm 1, \pm 2 $, whereas the Chern numbers of $\rm VB2$ remain at $C = \pm 1$ as shown in Fig.\ref{fig:ChernPhase}(a). It is important to note that with the variation in $M/t_2$ and $\phi$, the $\rm VB1$ in AB stacked $\rm G/\tilde{G}$ bilayer shows multiple topological phase transitions. 
Furthermore, from Section.\ref{Sec:Bands}, it is observed that the Chern numbers are sensitive to the choice of $t_2$ values. Here, we calculated the Chern number variation for the parameter space of $t_2$ and $-\pi \leq \phi \leq \pi$. We also obtained the Chern phase diagram for $\rm \tilde{G}/\tilde{G}$ bilayer presented in Fig.\ref{fig:ChernPhase}(b). In this case, the Chern number varies between $C = 0, \pm 1, \pm 2, 3 $ in the $M_l/t_{2l}$ and $\phi_l$ parametric space considering $M_u=0, t_{2u}=0.1t_g$ and $\phi_u=\pi/2$. Moreover, we present the case with the upper and lower layers having different values of Haldane terms $t_2$ and $\phi$ in the $\rm \tilde{G}/\tilde{G}$ bilayers and thereby their key role on the bands topology. The non-trivial bands show multiple topological phase transition for AA and AB stacked $\rm \tilde{G}/\tilde{G}$ bilayer with the variation of lower Haldane flux ($\phi_l$) and upper Haldane flux ($\phi_u$) considering $M_l=M_u=0$, and $t_{2l}=t_{2u}=0.1t_g$ as shown in Fig.\ref{fig:chern_phi_l_phi_u}. But interestingly, the case of two layers with Haldane terms of opposite fluxes (i.e., $\phi_l=-\phi_u$) with the same magnitude preserve the system’s overall symmetry, rendering the bands to be trivial.
\begin{figure}
    \centering
    \includegraphics[width=\columnwidth]{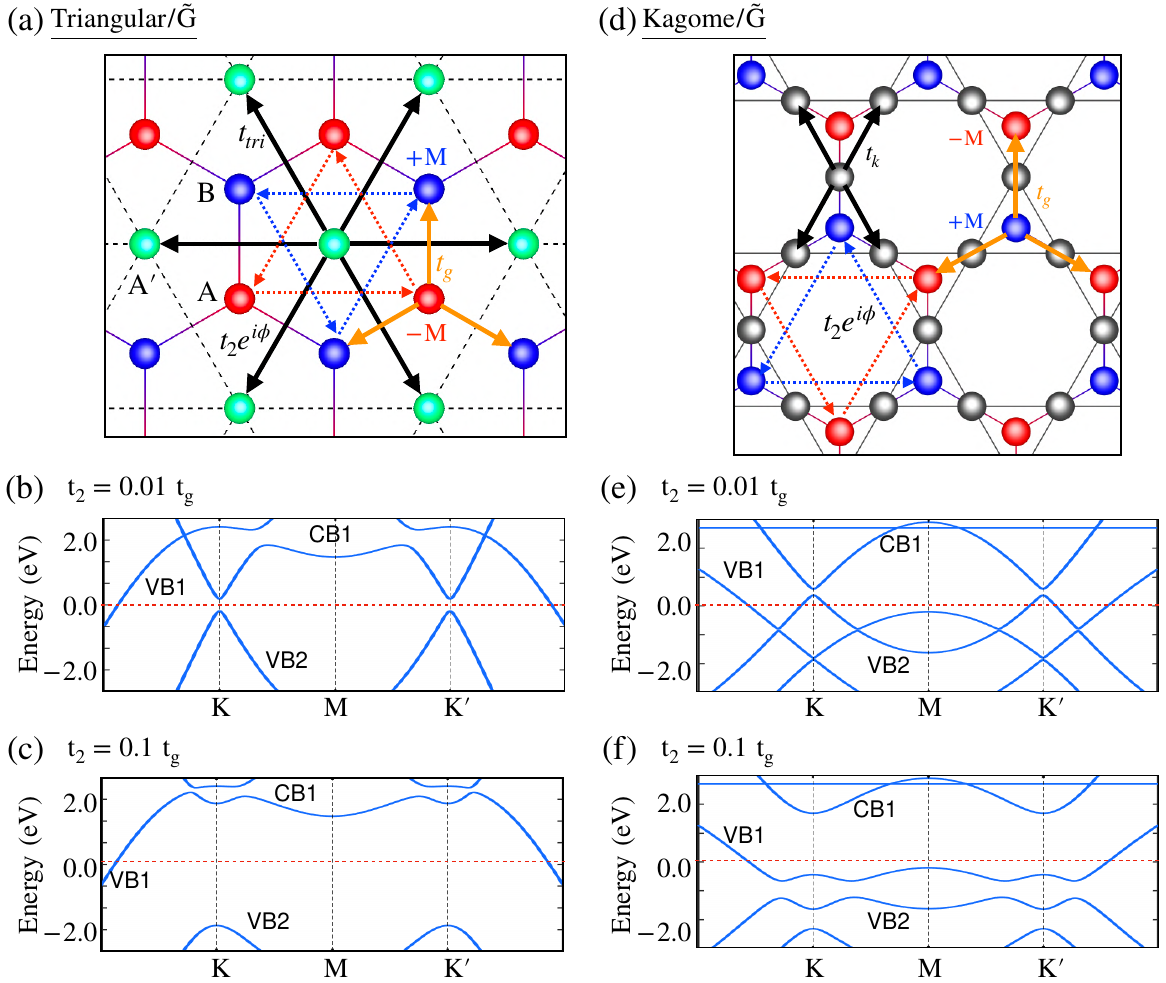}
    \caption{(a) and (d) Schematic diagram of the triangular/Haldane ($\rm T/\tilde{G}$) and Kagome/Haldane ($\rm K/\tilde{G}$) bilayers from the top view. The Haldane layer ($\tilde{G}$) Hamiltonian is defined in Sec.(\ref{TB_hamil}) and Fig.\ref{fig:schem}. In $\rm T/\tilde{G}$ and $\rm K/\tilde{G}$, we consider only one sublattice, for the triangular lattice with an NN hopping strength $t_{k}$. (b),(c),(e) and (f) represent the energy spectrum of $\rm T/\tilde{G}$ and $\rm K/\tilde{G}$ bilayer for two different NNN hopping strength as $t_2=0.01t_g$ and $t_2=0.1t_g$.}
    \label{hetero_bands}
\end{figure}
\begin{figure}
    \centering
    \includegraphics[width=\columnwidth]{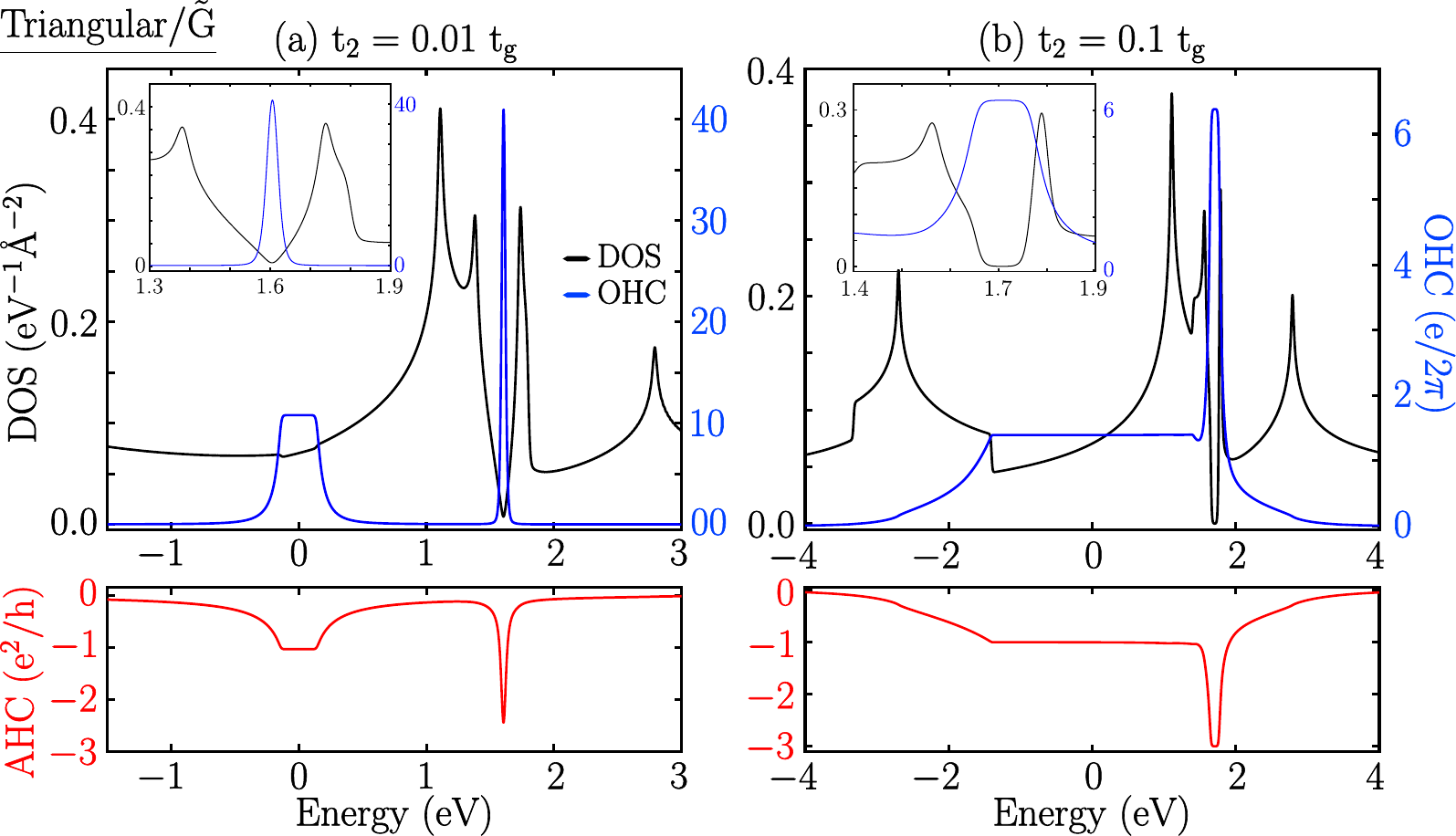}
    \includegraphics[width=\columnwidth]{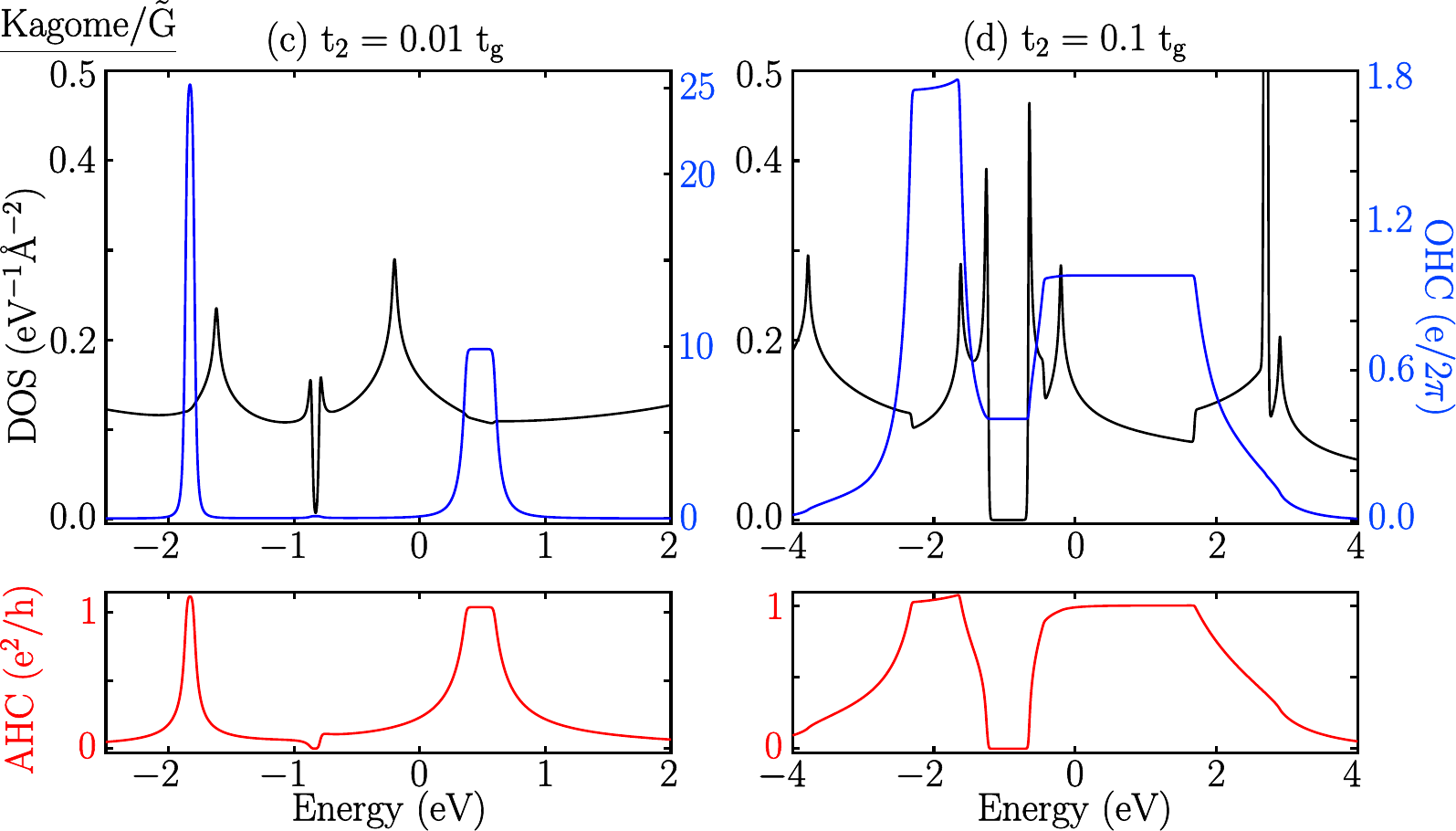}
    \caption{Orbital Hall conductivity of $\rm T/\tilde{G}$ and $\rm K/\tilde{G}$  bilayer for two different NNN hopping strength as $t_2=0.01t_g$ and $t_2=0.1t_g$ compared along with the AHC and DOS.}
    \label{hetero_ohc}
\end{figure}
\section{Heterostructures} \label{app.Hetero}
In heterobilayers consisting of dissimilar lattices, where one layer is of the Haldane type, our findings suggest the emergence of quantum orbital Hall phases. We investigated the Triangular/Graphene $\rm (T/\tilde{G})$ and Kagome/Graphene $\rm (K/\tilde{G})$ bilayers,   where the Haldane layer utilizes the Hamiltonian defined in Sec.(\ref{TB_hamil}). Notably, these heterobilayers also exhibit various Quantum Orbital Hall phases and can be tuned into orbital chern insulators by adjusting the next nearest neighbor (NNN) hopping strength. 
\subsubsection{Triangular/Graphene $\rm (T/\tilde{G})$} \label{app.Hetero.tg}
Here, we constructed a triangular layer on top of the Haldane-type graphene layer, where each lattice atom is exactly at the middle of the hexagon of $\tilde{G}$ with a triangular network, as shown in Fig.\ref{hetero_bands}(a). Hamiltonian of the triangular layer is 
\begin{equation}
   H_{tri}^u(\mathbf{k}) =2t_{tri}\sum_{i=1}^3 cos(\mathbf{k\cdot a_i})
\end{equation}
where hopping strength $t_{tri}=0.2t_g$, $t_g=-2.7 eV$. The coupling matrix between triangular layer and Haldane layer is 
\begin{equation}
T_{HT}= \frac{t_p}{t_g}\begin{bmatrix}
         d^l_x-i d^l_y \\ d^l_x+id^l_y
    \end{bmatrix}
\end{equation}
where the inter-layer hopping strength was $t_p=0.1t_g$. Because the \textit{TR} symmetry is broken with the $t_2e^{\pm i\phi}$ from the $\tilde{G}$, the $\rm T/\tilde{G}$ bilayer exhibits non-trivial bands with integer Chern numbers. The corresponding energy spectrum is presented in Fig.\ref{hetero_bands}(b) and (c) for two different values of $t_2=0.01t_g$ and $t_2=0.1t_g$, respectively. The bands of the $\rm T/\tilde{G}$ bilayer at the two Dirac points ($K$ and $K'$) have a conical shape with a gap. Moreover, with a large value of $t_2 = 0.1t_g$, the energy gaps are enhanced and all the bands are well separated. The calculated Chern numbers of $\rm VB1$ and $\rm VB2$ are $C=-2$ and $C=-1$, respectively, whereas the conduction band has Chern number $C=3$ for $t_2 = 0.01t_g$. These Chern numbers remained unaltered with increasing value of $t_2 = 0.1t_g$.  Furthermore, we calculated the orbital Hall conductivity of the $\rm T/\tilde{G}$ bilayer and compared it with that of the AHC along with the Density of States (DOS), as shown in Fig.\ref{hetero_ohc}(a) and (b). For the $t_2 = 0.01t_g$, both the OHC and AHC are quantized in the gap at charge neutrality, indicating that the $\rm T/\tilde{G}$ bilayer is an {\it Orbital Chern Insulator}. However, owing to the narrow gap at higher energies in the conduction band, the OHC and AHC have peak values. Furthermore, with the increased value of $t_2=0.1t_g$, there is an enhancement of the gaps, and the quantized values of OHC and AHC have extended for the entire gap. Interestingly, the peak of OHC and AHC at higher energy in the conduction band gets quantized with $t_2=0.01t_g$ making it have Orbital Ferromagnetism in BZ. 
\subsubsection{Kagome/Graphene $\rm (K/\tilde{G})$} \label{app.Hetero.kg}
In the Kagome/Graphene ($\rm (K/\tilde{G})$) bilayer, we define the Kagome layer with three types of atoms, which are situated on top of the midpoint of the three C-C bonds of $\tilde{G}$ layer shown in Fig.\ref{hetero_bands}(d). The Hamiltonian of the Kagome layer can be written as
\begin{equation}
 H^{u}_{Kagome}(k)=\begin{bmatrix} H_{A'A'}(k) & H_{A'B'}(k) & H_{A'C'}(k)\\
cc & H_{B'B'}(k) & H_{B'C'}(k) \\
cc & cc        & H_{C'C'}(k)
\end{bmatrix}
\end{equation}
where $H_{A'A'}=H_{B'B'}=H_{C'C'}=0$, $H_{A'B'}=2t_k\cos{(\mathbf{k \cdot a_1/2})}$, $H_{A'C'}=2t_k\cos{(\mathbf{k \cdot a_2/2})} $ and $H_{B'C'}=2t_k\cos{(k \cdot a_3/2)}$ with $t_k=0.5 t_g$, $t_g=-2.7 eV$. The coupling matrix between the two layers is
\begin{equation}
    T_{HK}= \begin{bmatrix}
        H_{AA'} & H_{AB'} & H_{AC'}\\
        H_{BA'} & H_{BB'} & H_{BC'}
    \end{bmatrix}
\end{equation}
where $H_{AA'}=t_{p}e^{i\mathbf{k \cdot e_1/2}}$,$H_{AB'}=t_{p}e^{i\mathbf{k \cdot e_3/2}}$, $H_{AC'}=t_{p}e^{i\mathbf{k \cdot e_2/2}}$ and $H_{BA'}=H_{AA'}^{*}$,$H_{BB'}=H_{AB'}^{*}$,$H_{BC'}=H_{AC'}^{*}$, where the parameters $t_p=0.2t_g$, which is inter-layer coupling strength stronger than the $\rm T/\tilde{G}$ bilayer. In the case of the Kagome/Haldane ($\rm K/\tilde{G}$) bilayer, the low energy bands show a gap opening just above the charge neutrality at the Dirac points ($K$ and $K'$) with the $t_2=0.01t_g$ and $t_p=0.2t_g$ as shown in Figs.\ref{hetero_bands} (e) and (f), respectively. The gaps between the bands were found to be enhanced with increasing value of $t_2=0.1t_g$, and all the bands near the charge neutrality are isolated with a less dispersed nature. The Chern numbers of isolated bands $\rm VB1,VB2$ and $\rm CB1$ are $C=1,C=-1$ and $C=1$ respectively, and are unaltered with increasing value of $t_2=0.1t_g$. The calculated OHC and AHC along with the DOS, are compared in Fig.\ref{hetero_ohc} (c) and (d). In this case also, the OHC along with the AHC is quantized in the gaps for both $t_2=0.01t_g ~\& ~0.1tg$, in conduction band. There is a peak in the gap between $\rm VB1$ and $\rm VB2$ in the valence band, and OHC and AHC are quantized for an increased value of $t_2=0.1tg$. Interestingly, the OHC was quantized alone in the gaps between $\rm VB1$ and $\rm CB1$ with an increasing value of $t_2=0.1tg$. A $\rm K/\tilde{G}$ bilayer with appropriate band filling and $t_2$ tuning can exhibit orbital ferromagnetism and quantum orbital Hall phase. 
\bibliography{sample}

\end{document}